\documentclass[twocolumn]{aastex63}

\let\textbf\relax

\received{June 1, 2019}
\revised{January 10, 2019}
\accepted{\today}

\submitjournal{ApJ}

\shorttitle{Interaction of an S-star with the environment of Sgr~A*}
\shortauthors{Pei$\beta$ker et al.}

\graphicspath{{./}{figures/}}

\newcommand{\at}{\makeatletter @\makeatother}

\begin{document}

\title{First observed interaction of the circumstellar envelope of an S-star with the environment of Sgr~A*}

\correspondingauthor{Florian Pei{\ss}ker}
\email{Peissker@ph1.uni-koeln.de}

\author[0000-0002-9850-2708]{Florian Pei$\beta$ker}
\affil{I.Physikalisches Institut der Universit\"at zu K\"oln, Z\"ulpicher Str. 77, 50937 K\"oln, Germany}

\author[0000-0002-5728-4054]{Basel Ali}
\affil{I.Physikalisches Institut der Universit\"at zu K\"oln, Z\"ulpicher Str. 77, 50937 K\"oln, Germany}

\author[0000-0001-6450-1187]{Michal Zaja\v{c}ek}
\affil{Center for Theoretical Physics, Al. Lotników 32/46, 02-668 Warsaw, Poland}
\affil{I.Physikalisches Institut der Universit\"at zu K\"oln, Z\"ulpicher Str. 77, 50937 K\"oln, Germany}
\affil{Max-Plank-Institut f\"ur Radioastronomie, Auf dem H\"ugel 69, 53121 Bonn, Germany}

\author[0000-0001-6049-3132]{Andreas Eckart}
\affil{I.Physikalisches Institut der Universit\"at zu K\"oln, Z\"ulpicher Str. 77, 50937 K\"oln, Germany}
\affil{Max-Plank-Institut f\"ur Radioastronomie, Auf dem H\"ugel 69, 53121 Bonn, Germany}

\author[0000-0002-3004-6208]{S. Elaheh Hosseini}
\affil{I.Physikalisches Institut der Universit\"at zu K\"oln, Z\"ulpicher Str. 77, 50937 K\"oln, Germany}
\affil{Max-Plank-Institut f\"ur Radioastronomie, Auf dem H\"ugel 69, 53121 Bonn, Germany}

\author[0000-0002-5760-0459]{Vladim\'ir Karas}
\affil{Astronomical Institute, Czech Academy of Sciences, Bo\v{c}n\'{i} II 1401, CZ-14100 Prague, Czech Republic}

\author[0000-0002-8382-2020]{Yann Cl\'enet}
\affil{LESIA, Observatoire de Paris, Universit\'e PSL, CNRS, Sorbonne Universit\'e, Universit\'e de Paris, 5 place Jules Janssen, 92195 Meudon, France}

\author{Nadeen B. Sabha}
\affil{Institut f\"ur Astro- und Teilchenphysik, Universit\"at Innsbruck, Technikerstr. 25, 6020 Innsbruck, Austria}

\author{Lucas Labadie}
\affil{I.Physikalisches Institut der Universit\"at zu K\"oln, Z\"ulpicher Str. 77, 50937 K\"oln, Germany}

\author{Matthias Subroweit}
\affil{I.Physikalisches Institut der Universit\"at zu K\"oln, Z\"ulpicher Str. 77, 50937 K\"oln, Germany}

\begin{abstract}

Several publications highlight the importance of the observations of \textbf{bow shocks} to learn more about the surrounding \textbf{interstellar medium and} radiation field. We \textbf{revisit} the most prominent dusty and gaseous \textbf{bow shock} source, X7, close to the supermassive black hole, Sgr~A*, \textbf{using multiwavelength analysis}. For \textbf{the purpose of this study}, we use SINFONI (H+K-band) and NACO ($L'$- and $M'$-band) data-sets between 2002 and 2018 \textbf{with additional COMIC/ADONIS+RASOIR ($L'$-band) data of 1999}.
By analyzing the line maps of SINFONI, we identify a velocity of  $\sim 200$ km/s from the tip to the tail. Furthermore, \textbf{a combination} of the \textbf{multiwavelength} data of NACO and SINFONI in the $H$-, $K$-, $L'$-, and $M'$-band results in a \textbf{two-component} black-body fit that {implies that} X7 is a dust-enshrouded stellar object.
The observed ongoing elongation and orientation of X7 in the Br$\gamma$ \textbf{line maps} and the NACO $L'$-band continuum indicate a wind arising at the position of Sgr~A* or at the IRS16 complex. Observations after 2010 show that the dust and the gas shell seems to be decoupled in projection from its stellar source S50. The data also implies that the tail of X7 gets thermally heated up \textbf{due to the presence of S50}. The gas emission at the tip is excited because of the related forward scattering (Mie-scattering)\textbf{, which will continue to influence the shape of X7 in the near future}. 
\textbf{In addition}, we \textbf{find excited} [FeIII] lines, which underline together with the recently \textbf{analyzed} dusty sources and the Br$\gamma$-bar the uniqueness \textbf{of this source.}

\end{abstract}

\keywords{editorials, notices --- 
miscellaneous --- catalogs --- surveys}

\section{Introduction}

In the center of our Galaxy, \textbf{the prominent} variable radio source Sgr~A* \textbf{is located} \citep{Balick1974}. \textbf{This source} emits across a broad range of wavelengths, ranging from the radio up to the X-ray domain, with the peak at submillimeter wavelengths \citep[see e.g.][and references therein]{2010RvMP...82.3121G,Eckart2017}. Although Sgr~A* is a low-luminosity source, its monitoring has been of \textbf{high} interest because of order-of-magnitude flares in the near-infrared and X-ray domains \citep{Witzel2012, Do2019}. Because of its \textbf{nonthermal} radiative properties, compact nature, variability, and \textbf{position} at the Galactic center, it has been associated with a supermassive black hole \textbf{(SMBH)} since its discovery \citep{Lynden-Bell1971}, with most of the alternatives being ruled out based on the current observational data \citep{Eckart2017}.

Sgr~A* is also the only \textbf{SMBH to date}, where we can detect and monitor \textbf{orbiting stars. Some of them are located inside the S-cluster, hence, they are called S-stars. These stars show pericentre} distances of \textbf{several $100$ AU} \citep{Gillessen2009, Parsa2017, Ali2020}. \textbf{Recently} discovered \textbf{stars push this distance an} order of magnitude closer \textbf{to the SMBH} \citep{Peissker2020a, Peissker2020b}. These \textbf{S-stars are widely covered by many publications}. For example, \textbf{\cite{Eckart1996Natur}} derived from the stellar proper motion \textbf{a direct} mass estimate of Sgr~A*. \textbf{In addition}, \cite{Ghez2002} and \cite{Eckart2002} found stellar accelerations based on the orbital curvature. \cite{Genzel2000} derived a velocity dispersion as a function of the distance of \textbf{S-stars} and \textbf{found values of up to several hundred km/s.}\newline

%Also, \cite{Alexander2003} proposed the existence of around 120 of the so called squeezars. These faint and fast moving stars fill up the region close to Sgr~A* which is expected to be statistically empty based on the predictions of power-law stellar density profiles \citep{2006ApJ...645.1152H,Zajacek2018}. \citet{Zajacek2018} investigated the probability of occurrence of these possible stars which are ideal probes of strong-gravity relativistic effects. In \cite{Peissker2020d}, we find that S4711 is a perfect squeezar candidate that is in line with the proposed properties like, e.g., $r_p\,<\,120\,AU$ by \cite{Alexander2003}.

%Besides the \textbf{presence} of the fast moving S-stars and their at least partially relativistic orbits\textbf{no need for this sentence} \citep[see e.g.,][]{Parsa2017, GravityCollaboration2020}, other objects with properties that exceed the description of a star\textbf{this sentence in unclear to me}, can also be found in the S-cluster\textbf{the S-cluster is not a region, can be found in the inner or central arcseconds}. 

One of the \textbf{controversial but also interesting} sources in the field of view (FOV) is the Galactic center (GC) gas cloud G2 \citep{Gillessen2012, EckartAA2013,Valencia-S.2015, Shahzamanian2017, Zajacek2017, Peissker2020c} also known as the Dusty S-cluster Object (DSO)\footnote{Since the nature of this source is better represented by the name DSO, we will use this throughout the manuscript.}. This object was found on its way \textbf{approaching Sgr A*} in the Doppler shifted Br$\gamma$ maps of SINFONI, a near-infrared (NIR) instrument mounted at the Very Large Telescope (VLT, Chile/Paranal). In combination with the observed dust emission in the $L'$-band ($3.8\,\mu m$) with NACO (also operating in the NIR, mounted at the VLT), the authors of \cite{Gillessen2012, Gillessen2013, Pfuhl2015} claimed that the object will get disrupted during or after its periapse passage. Later on, \cite{Plewa2017} stated that the density of the ambient medium of Sgr~A* is too low to cause a disruptive event. Even more, they excluded the possibility of a drag force acting on the DSO. In contrast, \cite{Gillessen2019} reported a drag force that \textbf{influenced} the observed Doppler shifted Br$\gamma$ line shape. This underlines the ongoing confusion about the nature of the source. However, in \cite{Peissker2020c} we present a spectral energy distribution (SED) derived from the H-, K-, and $L'$-band data of NACO and SINFONI. This SED \textbf{consists of a dusty and stellar component} and shows that the DSO is more likely a Young Stellar Object instead of a coreless $\sim\,3\times M_{\oplus}$ cloud that moves on a Keplerian orbit around a $4.1\times 10^6 M_{\odot}$ super massive black hole.

%\textbf{In my opinion, this subsection is irrelevant to this study and this argument can be presented when the confusion about the nature of this object is cleared, i.e., when there is new observational data that is worth publication and not already discussed debate. This will shift the mind of the reader to a different topic and not the current presented source.}

\cite{Clenet2003a} and \cite{Clenet2005a} reported for the first time two comet-shaped sources, namely X3 and X7. These dusty objects can be found in the mid-infrared (MIR) but also show a NIR \textbf{counterpart. Because of its close projected distance to X7, another line emitting source is located that we call X7.1 \citep[G5 in][]{Ciurlo2020}}.\newline
The identification of these objects is still challenging, which is \textbf{manifested} in Fig. 1 in \textbf{\cite{Peissker2020c}. The potentially} temporary distance of X7 and X7.1/G5 can lead to confusion about the identification without \textbf{spectroscopic} analysis.
It is, for instance, not clear why the dusty object X7.1/G5 with an approximate L-band magnitude of 14.11 mag ($\sim 0.57\,{\rm mJy}$) can neither be observed in the NACO ($L'$-band) data presented in this work nor in the shown $3.8\,\mu m$ continuum data in \cite{Ciurlo2020} (see extended data Fig. 2 in the related publication). A dust-enshrouded source with a stellar counterpart should be detectable in the $L'$-band as presented in \cite{Peissker2020c}. A reliable approach is the spectroscopic analysis in combination with \textbf{multiwavelength} observations. This underlines the need for broad observation programs. \textbf{Following} the example of the DSO \textbf{and X7.1}, we \textbf{emphasize a multiwavelength} analysis of these \textbf{(potentially)} dust-enshrouded stars. \textbf{With} the observational coverage of different bands in combination with spectroscopy\textbf{, the confusion about the nature of these objects can be minimized} \citep[see][]{Zajacek2017}. 

However, \cite{muzic2010} analysed the X7 source in detail and showed the connection to a possible nuclear wind that arises at the position of Sgr~A*. This wind t\textbf{is also mentioned in} several observational and theoretical publications \citep{muzic2007, Zajacek2016, Yusef-Zadeh2017-ALMAVLA, Peissker2020c, Peissker2020d, Yusef-Zadeh2020}.
\textbf{In this regard,} \citet{Peissker2019} reported a new \textbf{bow shock} source \textbf{in the central arcsecond} that the authors call X8 \citep[G6 in][]{Ciurlo2020} because of its close \textbf{projected} distance to X7. These two objects are the closest \textbf{bow shock} sources that could be used to determine the properties of a possible wind that arises at the position of Sgr~A*.

In this work, we will update the analysis of X7 done by \citet{muzic2007, muzic2010} with the help of SINFONI \textbf{integral field} spectroscopy and NACO continuum data that cover almost 16 years. Additionally, we use $L'$-band continuum COMIC/ADONIS+RASOIR data of 1999 to extend the analysis of X7 to about 20 years. This work is part of \textbf{a broader} investigation that is split up in two publications. Here, we emphasize the observational results and give an outlook on the second part where we theoretically investigate the observed source X7. In the second part of this survey,  we will apply two models to describe an open and \textbf{closed} \textbf{bow shock} based on the work of \citet{Wilkin1996, Wilkin2000} and \citet{Christie2016}. 
The spectroscopic capabilities of SINFONI give us an access to investigate the \textbf{velocity along} the \textbf{bow shock} source that could help to describe the nuclear and the stellar wind interaction as well as prominent Doppler-shifted emission lines. Furthermore, we will investigate the close \textbf{projected} distance of S50 to X7. This S-cluster star can be associated with the stellar counterpart of X7 and seems to interact with the dust tail of the \textbf{bow shock} source. \textbf{In the} \textbf{multiwavelength} analysis, we also model a two-component SED of X7.
We also witness the ongoing decoupling in projection of the dusty and gaseous shell of S50 that is associated with X7. %We \textbf{offer} different interpretations \textbf{such as} dust waves and a tidal interaction with Sgr~A* \textbf{that} will be discussed \textbf{in Sec. \ref{sec:DiscussionConclussion}}.

\textbf{In the following Sec. \ref{sec:2}, we introduce the used instruments and the analyzing techniques. The results of the analysis are presented in Sec. \ref{sec:3}. Section \ref{sec:DiscussionConclussion} summarizes the results and provides an outlook for future observations. In Appendix \ref{sec:appendix}, we will give some supplementary information regarding the analysis and a possible scenario. In addition, we list the SINFONI and NACO data that were used for the analysis.}

\section{Data \& Analysis}
\label{sec:2}
In this section, we will give a brief overview about the used instruments, the data reduction, and the applied analysis tools.

\subsection{SINFONI \& NACO}

The Spectrograph for Integral Field Observations in the Near Infrared (SINFONI) was mounted on the VLT and undergoes an upgrade \citep[for further information about the upgrade, see][]{Eris_1_2014, Eris_2_2014, Eris_3_2014}. SINFONI operates in the NIR and provides observations in the J- ($1.10\,-\,1.40\,\mu m$), H- ($1.45\,-\,1.85\,\mu m$), K- ($1.95\,-\,2.45\,\mu m$), and H+K-band ($1.45\,-\,2.45\,\mu m$). The output files of the ESO pipeline are in the shape of a 3 dimensional \textbf{data cube}. \textbf{This data cube} consists of 2 spatial dimensions and 1 spectral dimension. The components of the \textbf{data cubes} are described in \textbf{spaxels \citep[pixels containing a spectrum, see][]{hoertner2012}} rather than \textbf{pixels}. With SINFONI, we are able to isolate single emission lines in the H+K-band \textbf{to} create channel (line) maps. In comparison, the NACO\footnote{Nasmyth Adaptive Optics System (NAOS) \& Near-Infrared Imager and Spectrograph (CONICA) = NACO} instrument works also in the J, $L'$, and $M'$-band \citep{Lenzen2003, Rousset2003}. Since dust can be traced in higher wavelengths, the $L'$-band setup of NACO is favored for the search of the dusty \textbf{bow shock} source. \newline
In both cases, we apply the usual \textbf{data reduction} steps like, e.g., dark- and flat-field corrections. We also apply the mandatory sky correction to the adaptive optics (AO) corrected data. Additional correction steps are described in detail in \cite{Peissker2019, Peissker2020a, Peissker2020c, Peissker2020d, Peissker2020b} where the here analyzed data is also used. Please consider also the Appendix \ref{sec:appendix_Data} for a detailed overview about the used data. \newline
We also note that a \textbf{part} of this data was used in \cite{Parsa2017}. The authors describe the Schwarzschild precision of S2, which was independently confirmed by \cite{GravityCollaboration2020}\textbf{. The} collaboration used GRAVITY, an interferometric instrument with a resolution almost one magnitude better than NACO\footnote{\cite{GravityCollaboration2020} \textbf{state} 3 mas compared to 27 mas of the $L'$-band setting of NACO.}. This underlines the robustness and validity of the NACO data.

\subsection{COMIC/ADONIS+RASOIR}

The NIR camera COMIC was installed in La Silla/Chile at the 3.6 m telescope and used the AO system ADONIS+RASOIR \citep{Beuzit1994, Lacombe1998}. It operated in the J-, H-, K-, $K'$-,  $L'$-, and M-band with two different plate scales (35 mas/pixel and 100 mas/pixel). It was optimized for L- and M-band observations and decommissioned in 2001 \citep{Pasqunini1996}. For the here presented data, the exposure time was set to 10 seconds. The observational pattern was \textbf{chosen} to be s-o-s (sky-object-sky) followed by flat and dark exposures. For combining the data, we use the shift and add algorithm to maximize the \textbf{signal-to-noise (S/N)} ratio. This is followed by rebining the data \textbf{to} smooth sharp edges caused by the resolution. The COMIC/ADONIS+RASOIR data analyzed in this work was first published in \cite{clenet2001}.

\subsection{High-pass filter}

\textbf{Depending on the scientific goal,} a suitable frequency \textbf{pass filter} can improve the amount of accessible image information. %We already used high-pass filter in \cite{Peissker2019, Peissker2020a, Peissker2020c}. 
In the case of an elongated source like X7, using the Lucy-Richardson algorithm \citep{Lucy1974} is not the most satisfying option for the $L'$-band NACO data. However, using a high-pass filter like the smooth subtract algorithm can reveal the stellar component of the \textbf{bow shock} source in the K-band SINFONI data if the object is confused with nearby S-stars. For that, we are subtracting a smoothed version of the input image file. The size of the Gaussian that is used for the \textbf{smoothing} should be of the order of the related image point spread function (PSF). The resulting smooth subtracted image should then be \textbf{resmoothed} with a Gaussian PSF that can be $10-20\%$ smaller than the image PSF. With this technique, the influence of overlapping PSF-wings can be minimized.

\section{Results}
\label{sec:3}
This section shows the results of the survey of the X7/S50-system between 2002 and 2018. We present the \textbf{line map} and continuum detection of the \textbf{bow shock} source X7 and show the ongoing implied decoupling of the shell from its central star, S50. Furthermore, we compare the observations with \textbf{the} published COMIC/ADONIS+RASOIR data of 1999 and apply a photometric analysis to the NACO images.

\subsection{Line map and velocity gradient detection of X7}

Throughout the  SINFONI data between 2005 and 2018, the source X7 can be at least partially observed. A key parameter is the FOV. \textbf{Hence,} the SINFONI data in 2006, 2008, 2014, 2015, and 2018 can be used for a \textbf{detailed} analysis.
\begin{figure*}[htbp!]
	\centering
	\includegraphics[width=1.\textwidth]{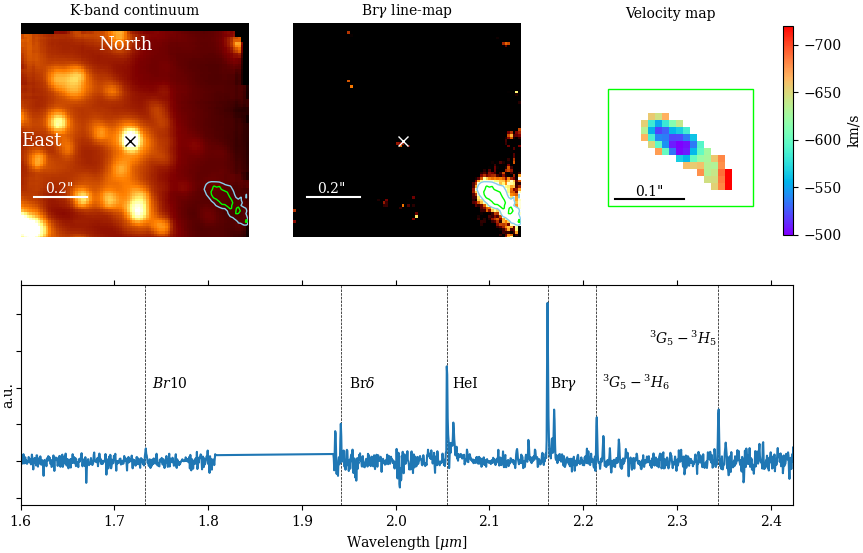}
	\caption{\textbf{Bow shock} source X7 in 2018. The upper left plot shows the K-band continuum with SINFONI of the S-cluster where the black $\times$ marks the position of Sgr~A*. The marked position of the SMBH coincides with the S-cluster star S2 in 2018 because of its pericenter passage. We adapt the \textbf{line map} contours of X7 from the Br$\gamma$-emission detection at $2.162\,\mu m$ (upper middle image) and include them in the continuum image. The upper right panel shows a zoomed-in map of the Br$\gamma$ \textbf{line map} with a spatial pixel size of 12.5 mas. We present the velocity in this panel based along the \textbf{line map} detection of X7 in the SINFONI cube of 2018. For that, we fit a Gaussian to the spectrum of every spaxel in order to create a confusion free velocity map. In the lower panel, the spectrum of X7 can be seen where we mark prominent lines. The related spectrum is integrated over all pixel shown in the top right panel \textbf{('Velocity map')}. The telluric emission between $1.80\,-\,1.93\,\mu m$ is clipped. The most prominent blue-shifted emission lines are Br$\delta\,$\at$\,1.9414\,\mu m$, HeI$\,$\at$\,2.0545\,\mu m$, Br$\gamma\,$\at$\,2.1619\,\mu m$, $[{\rm FeIII}] ^3G_5 - ^3H_6\,$\at$\,2.2144\,\mu m$, and $[{\rm FeIII}] ^3G_5 - ^3H_5\,$\at$\,2.2344\,\mu m$. Next to the blue-shifted HeI and Br$\gamma$ line, we observe a red-shifted emission that is related to X7.1/G5. This source is in projection spatially close to X7 \textbf{\citep{Peissker2020c}}.}
\label{fig:X7_overview_2018}
\end{figure*}
By \textbf{analyzing continuum} subtracted line maps, we find the length from the tip to the tail of the detected Doppler-shifted Br$\gamma$ emission to be around \textbf{0.23"} in 2006. Furthermore, we measure \textbf{the} length of the \textbf{bow shock} of about \textbf{0.35''} in 2018.\newline 
\begin{table*}[hbt!]
    \centering
    \begin{tabular}{cccc}
         \hline %& flux [10$^{-16}$ erg/s/cm$^{2}$]
         \hline
           Spectral line ($@$rest wavelength [$\mu$m]) & transition & central wavelength [$\mu$m] & Velocity [km/s] \\
         \hline
         
          Br10 $@$1.7366 $\mu$m &         n=10-4         & 1.7335 $\pm$ 0.0002 & -536 \\
          Br$\delta$ $@$1.9450 $\mu$m &   n=8-4          & 1.9414 $\pm$ 0.0002 & -555 \\
          HeI $@$2.0586 $\mu$m &    $2p^1 P^0-2s^1S$     & 2.0545 $\pm$ 0.0002 & -597 \\
          Br$\gamma$ $@$2.1661 $\mu$m & n=7-4            & 2.1619 $\pm$ 0.0002 & -582 \\
          $[FeIII]@$2.1451 $\mu$m & $^3G_3\,-\,^3H_4$    & 2.1414 $\pm$ 0.0002 & -517 \\
          $[FeIII]@$2.2178 $\mu$m & $^3G_5\,-\,^3H_6$    & 2.2145 $\pm$ 0.0002 & -446 \\
          $[FeIII]@$2.2420 $\mu$m & $^3G_4\,-\,^3H_4$    & 2.2379 $\pm$ 0.0002 & -549 \\
          $[FeIII]@$2.3479 $\mu$m & $^3G_5\,-\,^3H_5$    & 2.3444 $\pm$ 0.0002 & -447 \\
          H$_2@$2.4065 $\mu$m & v=1-0 Q(1)               & 2.4045 $\pm$ 0.0002 & -250 \\
          H$_2@$2.4134 $\mu$m & v=1-0 Q(2)               & 2.4130 $\pm$ 0.0002 & -50  \\
          H$_2@$2.4237 $\mu$m & v=1-0 Q(3)               & 2.420  $\pm$ 0.0002 & -458 \\
       \hline
    \end{tabular}
    \caption{Observed emission and absorption lines of X7/S50 in 2018. All emission lines are related to Fig. \ref{fig:X7_overview_2018} whereas the H$_2$ absorption lines are shown \textbf{for better} visibility in Appendix~\ref{appendix_h2_emission}, Fig. \ref{fig:H2_emission}. The typical uncertainty of the measured central wavelength (peak intensity) indicates the standard deviation \textbf{to} $2.5\,\times\,10^{-4}$. Hence, the uncertainty of the derived Doppler-shifted velocity is about $\pm\,35$ km/s.}
    \label{tab:emissionlines}
\end{table*}
Because of the high \textbf{S/N} ratio of the SINFONI data in 2018 (see Appendix \ref{sec:appendix_Data}) at the spatial position of the Doppler-shifted Br$\gamma$-emission of X7, we use this set to investigate the velocity gradient of the \textbf{bow shock} source.
For this purpose, we fit a Gaussian to the blue-shifted Br$\gamma$-line in the related spectral range ($2.16\pm0.04\,\mu m$). Afterwards, the related \textbf{spaxel} carrying the velocity information is copied to the same position in a new array that is as big as the input file. We manually mask the \textbf{close-by} source X7.1/G5 \citep[see][]{Ciurlo2020, Peissker2020c} and \textbf{non-linear} pixel.\newline 
In Fig. \ref{fig:X7_overview_2018}, the resulting velocity gradient is shown. We find a difference from the tip to the tail along the projected \textbf{bow shock} structure of $\approx 190\pm\,20$ km/s. Considering a spatial pixel scale of 12.5 mas in the H+K-band in the highest plate-scale setting of SINFONI and the measured projected \textbf{bow shock} length of 349 mas, we get a linear gradient of $\approx 7.1\pm\,1.0$ km/s/pixel in 2018. Furthermore, we find several prominent emission lines that indicate the presence of ionized gas (see Table \ref{tab:emissionlines}). \textbf{In several data sets, a H$_2$ emission triplet can be found (Appendix~\ref{appendix_h2_emission}, Fig. \ref{fig:H2_emission}). Due to crowding and therein the resulting possibility of confusion, we limit the analysis of the H$_2$ emission triplet to the data of 2018.}

\subsection{Continuum detection of X7}

\begin{figure*}[htbp!]
	\centering
	\includegraphics[width=1.\textwidth]{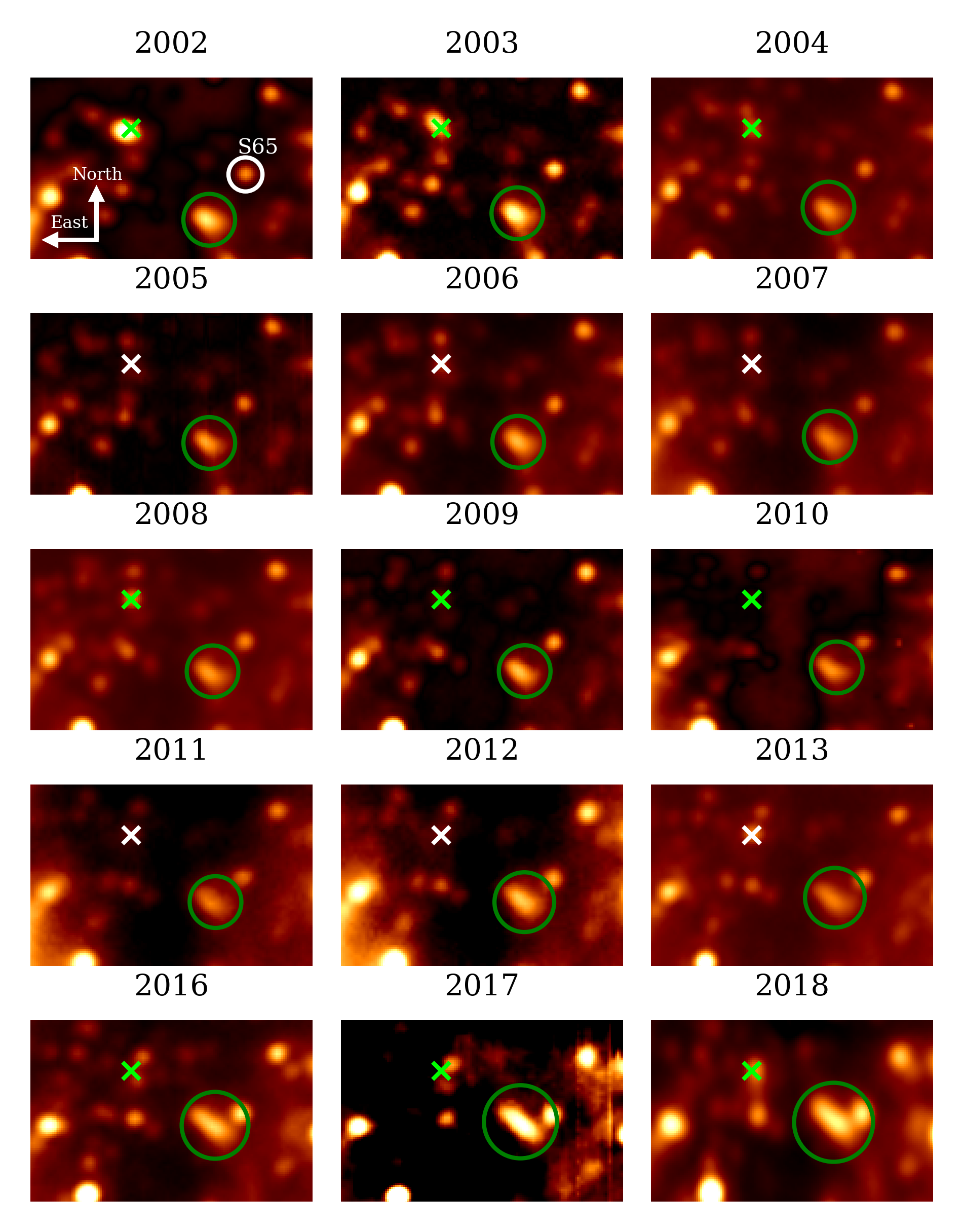}
	\caption{$L'$ images of the GC observed with NACO between 2002 and 2018. The size of every panel is $1.8"\,\times\,1.2"$. As indicated, north is up, east is to the left. The position of Sgr~A* is marked with a $\times$ and locked in every panel. The green circle indicates the position of the \textbf{bow shock} source X7. In the upper left panel, we show the position of the S-star S65 with a white circle. The related measured projected on-sky size and the position angle of X7 of every year are listed in Table \ref{tab:X7_length}.}
\label{fig:X7_2016_NACO}
\end{figure*}

$L'$-band \textbf{observation} of the \textbf{bow shock} source X7 in the close distance of the \textbf{S-cluster shows} that the object is always one of the most prominent sources in the close vicinity \textbf{of the SMBH} (see Fig. \ref{fig:X7_2016_NACO}). \textbf{The $L'$-band brightness and elongated shape of X7 underlines the unique character of the object.}\newline
After 2002, X7 \textbf{becomes increasingly} brighter than most of its nearby stars like, e.g., S1, S2, S61, and S71. The \textbf{bow shock} shape is clearly noticeable in the NACO $L'$-band (green circles indicate its position in Fig. \ref{fig:X7_2016_NACO}). After 2010, the source shows a more elongated shape with an approximate projected length of 333 mas in 2016. This is almost three times as much as the $L'$-band dust emission detected with NACO in 2002 (112 mas). Compared with the line emission area of the SINFONI data in 2006, 2008, 2009, 2013, 2014, 2015, and 2018, we find \textbf{matching} values for the gaseous emission (for a detailed list, see Table \ref{tab:X7_length}). Hence, the size of the projected area of the ionized gas is coinciding with the $L'$ continuum dust emission of X7. \newline
\begin{table*}[htb!]
\begin{center}
\begin{tabular}{cccc}\hline \hline
Year & $L'$ size (continuum) &  Br$\gamma$ size (line) & Position angle (projected)\\
 in [yr] & in [mas] & in [mas] & in [$^{\circ}$]\\  \hline
2002  &    112    &   -    & 42  \\
2003  &    148   &   -    &  44  \\
2004  &    162     &  -   &  45  \\
2005  &    180   &  -   &   45   \\
2006  &    200   &  230 &   45   \\
2007  &    206   &   -    &  45  \\
2008  &    229    &   230  & 45  \\
%\tabucline[1pt on 3pt]{1-4}
%\tabucline [1pt on 2.5pt off 2pt]{-}
\hline
2009  &    274   & 192   &  45  \\
2010  &    212  &  -  &    50    \\
\hline
2011  &    262  &  -    &   51   \\
2012  &    272   &  -     &  51  \\
2013  &    314   &   246   & 52  \\
2014  &     -     &  311   & - \\
2015  &     -     &  321  & - \\
2016  &    333    &  -    &  55    \\
2017  &    348     &    -  &  57   \\
2018  &    388     &  349  &  60*   \\
2019  &     -  &  -    &  - \\

\hline
\end{tabular}
\end{center}
\caption{Projected length of the \textbf{bow shock} source X7. The \textbf{line emission} length is extracted from the SINFONI \textbf{data cube} that shows the required FOV. The related \textbf{line map represents} the Doppler-shifted Br$\gamma$ emission line of the \textbf{bow shock}. From the NACO data, we derive the length of the \textbf{bow shock} from the $L'$-band continuum emission. Please note that the observation of X7 in 2009 can be distinguished in a  pre- (NACO, 2009.26) and post-event (SINFONI, 2009.47). We indicate the time of the pre- and post-event \textbf{(that shows a discontinuous behavior of the increasing elongation of the X7/S50 system)} with the horizontal lines before and after 2009 and 2010 respectively. To cover statistical variations, reading errors, background effects, and detector \textbf{irregularities,} we determine a spatial uncertainty of $\pm\,10$mas. For the position angle that is measured with respect to Sgr~A*, an uncertainty of $\pm\,2^{\circ}$ is given. The \textbf{asterisk} of the position angle measurement of 2018 \textbf{indicates $60^{\circ}$ as a lower limit. This lower limit is justified because} X7 is not \textbf{aligned} towards Sgr~A* \textbf{in 2018}.}
\label{tab:X7_length}
\end{table*}
Since we observe a clear increasing \textbf{projected size of X7} between 2002 and 2018, we investigate the $L'$-band data of 1999 to see if \textbf{this trend is also observable} in data before 2002. 
For \textbf{this purpose}, we use the investigated COMIC/ADONIS+RASOIR $L'$-band data in 1999 by \cite{clenet2001}. We apply a high-pass filter to reduce the influence of overlapping PSF. \textbf{Afterwards, we use a Gaussian that is about $50\%$ in size of the initial smoothing kernel (Appendix~\ref{subsec_COMIC_ADONIS}, Fig.~\ref{fig:comic_1999}) on the resulting high-pass filtered image}. In addition to some prominent members of the S-cluster, we identify at the expected position of S50 a \textbf{spherical} $L'$-band emission several magnitudes above the noise level. By comparing the closest NACO $L'$-band data to verify the COMIC/ADONIS+RASOIR identifications of 1999, we find matching positions for almost all stars/features.

\subsection{S50}

\cite{muzic2010} reported that the stellar counterpart of X7 could be associated with the S-cluster star S50. In Fig. \ref{fig:comic_1999} (right side), we present \textbf{the orbit} plots of S50 based on the analysis presented in \cite{Ali2020}. Throughout the available data covering the related spatial area, we find without confusion that the \textbf{bow shock} source X7 is moving along with S50 (see Fig. \ref{fig:turnover_point} and Appendix, Fig. \ref{fig:KL_ident_naco_2002}) till 2009.
S2 (K-band) and S65 (L-band) as the two brightest and therefore most prominent members of the S-cluster can always be observed in the same FOV as S50. Hence, we are using these two S-stars for a photometric analysis \textbf{to} investigate the magnitude of S50 and X7 in various bands (see Table \ref{tab:mag_S2_X7_S50}). In combination with \textbf{the published} SHARP data \citep{Schoedel2002}, we find a constant K-band magnitude of S50 of $m_K\,\approx\,16$ mag. \textbf{We find a similar magnitude with} NACO (VLT) data of 2007 and SINFONI (VLT) data of 2019. Based on the data that covers almost 20 years, we conclude that the S-cluster member S50 does not show a variable K-band emission. However, this is not the case for the $L'$-band continuum emission that seems to vary between 2008-2018. We will elaborate on this point in detail in Sec. \ref{sec:mag}. %\newline Based on the K-band magnitude, we are able to confidently derive the mass of S50 by adapting the method presented in \cite{Peissker2020d} using
%\begin{equation}
%    log\frac{M_S}{M_{\odot}}\,=\,km_K\,+\,b
%\end{equation}
%with $k\,=\,-0.1925$ and $b\,=\,3.885$. With this, we determine the mass of S50 to $M_S\,=\,6.3\,\pm\,1\,M_{\odot}$.

\subsection{Decoupling of X7 from S50?}

\begin{figure}[htbp!]
	\centering
	\includegraphics[width=.5\textwidth]{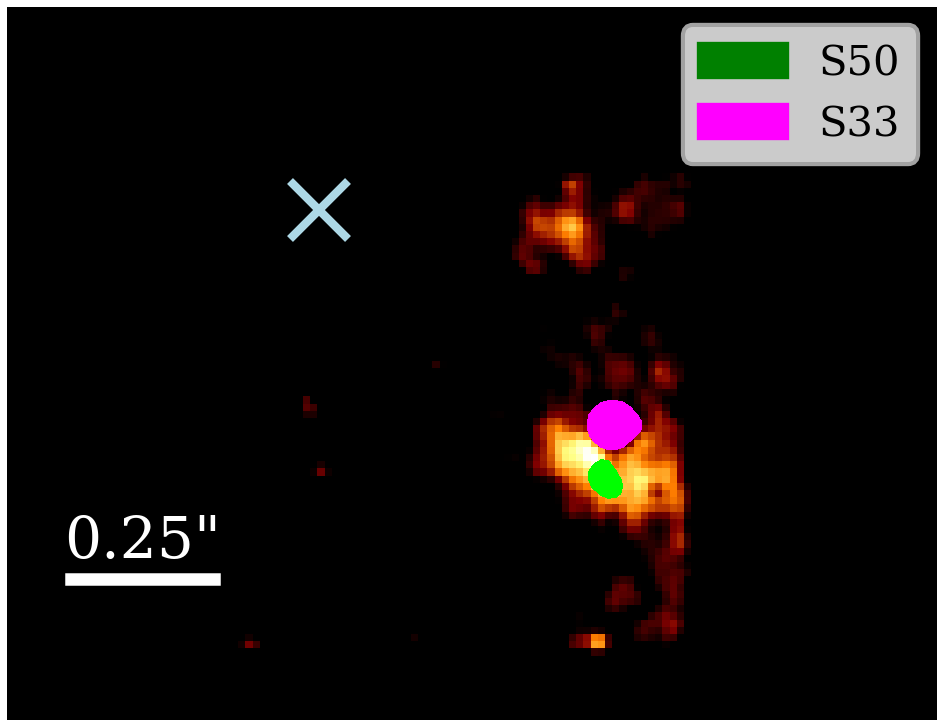}
	\caption{Doppler-shifted Br$\gamma$ \textbf{line map}s observed with SINFONI in 2009. East is to the left, north is up. The x marks the position of Sgr~A* which is derived by the offset of the well known orbit of S2. The filled contour line is related to the position of S50 and S33 (see the included legend). From the same \textbf{data cube}, we extract a K-band image ($2.0\mu m$-$2.2\mu m$) and isolate in the same wavelength window the Doppler-shift Br$\gamma$ line at around $2.161\mu m$.}
\label{fig:turnover_point}
\end{figure}

\begin{figure*}[htbp!]
	\centering
	\includegraphics[width=1.\textwidth]{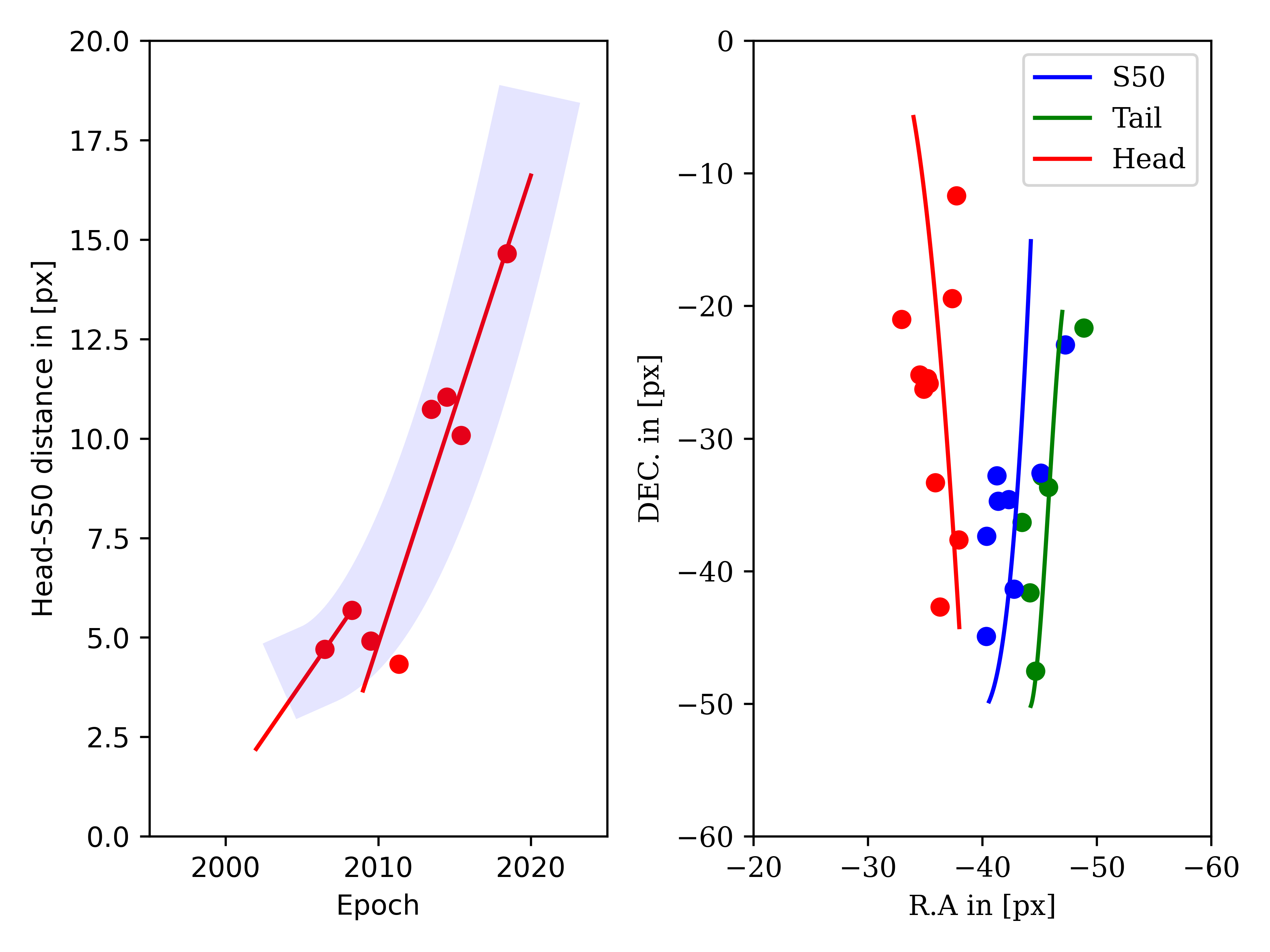}
	\caption{Distance of the head and tail of X7 in relation to the position of S50 and Sgr~A*. On the left, the distance of the head related to the position of S50 is plotted. In combination with Fig. \ref{fig:turnover_point}, we distinguish between two responsible processes for the evolution of the dust shell X7 which is reflected in the two fits. The overall trend is indicated with a blue transparent fit. On the right, the head (red), the tail (green), and S50 (blue) is shown with their position with respect to Sgr~A*. Again, the trend shows that the head is moving towards Sgr~A* and further away from S50. Typical uncertainties of about 1 px are not included to preserve the better readability of the plots. One pixel [px] corresponds to 12.5 mas.}
\label{fig:front_tip_distance}
\end{figure*}

\begin{figure*}[htbp!]
	\centering
	\includegraphics[width=1.\textwidth]{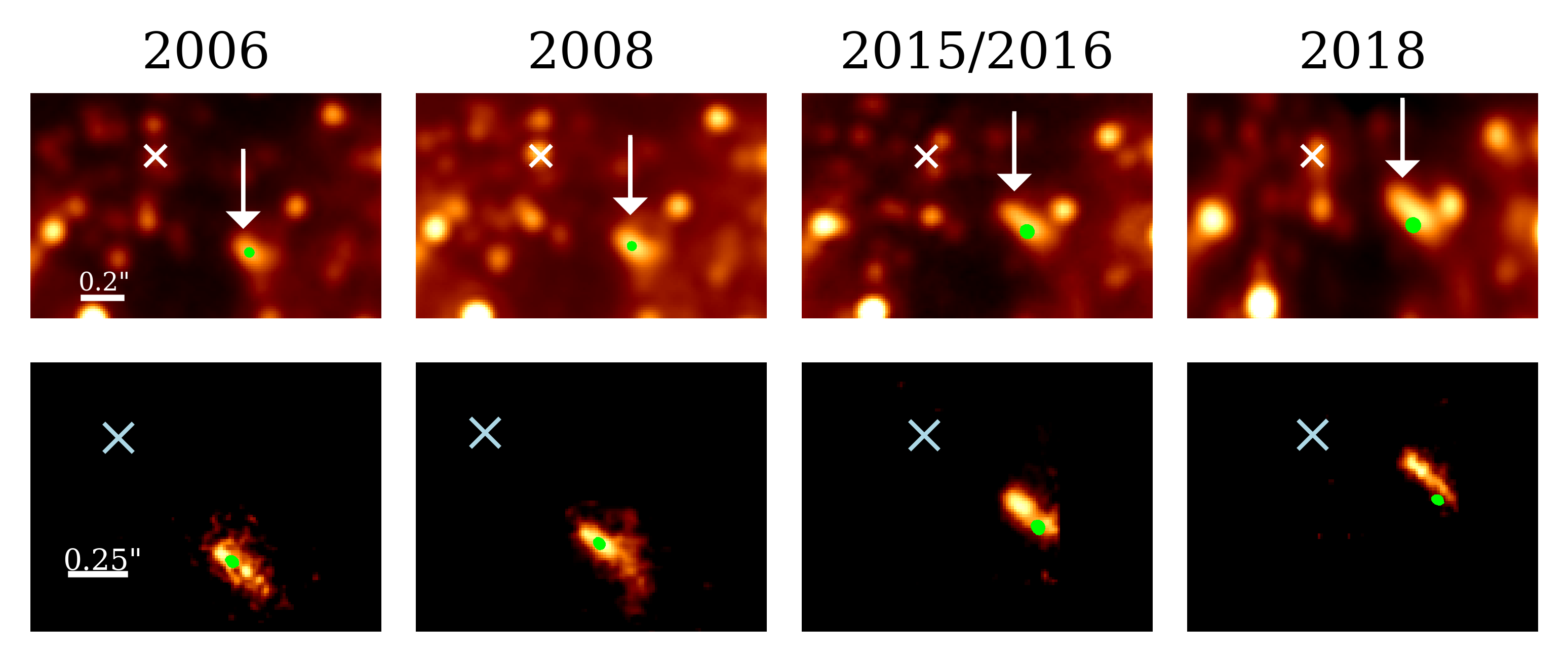}
	\caption{NACO $L'$-band continuum images (upper row) and SINFONI Doppler-shifted Br$\gamma$ line maps (lower row) displaying the immediate environment of Sgr~A*. Here we compare the appearance of the dust ($L'$-band) and the ionized gas (Br$\gamma$) in relation to the K-band of S50 which is indicated by the green dot. The green filled contour lines are extracted from the related K-band image of the same data (SINFONI) or year (NACO). Since NACO was decommissioned in 2014/2015, we use the $L'$- and K-band observations of early 2016 which are just 0.6 yr later than the displayed SINFONI \textbf{line map} of 2015. In every image, the colored $\times$ marks the position of Sgr~A* which is derived with the well observed S2 orbit. The pixel scale is identical in each row.}
\label{fig:distribution_gasdust_s50}
\end{figure*}

Based on the $L'$-band observations, we find a noticeable elongation of the source X7 that becomes \textbf{increasingly} prominent after 2009 (please see Fig. \ref{fig:X7_2016_NACO}). Comparing the NACO $L'$-band images with the SINFONI \textbf{line maps}, we find that the symmetric distribution of \textbf{gas} and dust cannot be observed \textbf{after 2009}. \textbf{Compared to the SINFONI data between 2006-2008 and 2010-2018}, the data shows a rather compact gas emission in 2009 (see Fig. \ref{fig:turnover_point}).
\textbf{Whereas the data of 2006-2008 shows a symmetrical gas-to-dust distribution with respect to S50 and X7}, we find that \textbf{this symmetry of the S50-X7 system is broken for the observations between 2010-2018}. Furthermore, we observe that the distance of the gaseous front of X7 (i.e., head) is increasing year-by-year \textbf{with respect to S50} (Fig. \ref{fig:front_tip_distance}). 
\textbf{In contrast, the back (or tail) of the Br$\gamma$ gas} emission does not show a comparable behavior \textbf{compared} to the head after 2009. \textbf{As previously described, this} leads to an asymmetric distribution of the gas \textbf{around the central stellar source} S50. \textbf{This broken symmetry between the shell and the star can also be observed in the NACO $L'$-band data} (see Fig. \ref{fig:distribution_gasdust_s50}).
\textbf{Hence, the data implies that the gas and dust shell starts} to detach in projection after 2010. This process can be tracked throughout the available NACO and SINFONI data beginning in 2009 and is indicated in Fig. \ref{fig:front_tip_distance}. \textbf{Furthermore, we find that the intensity maximum of the dust is located at a distance of less than 13 mas to the position of S50 (Fig. \ref{fig:X7_2016_NACO}). As a result, the tail of X7 gets increasingly brighter when comparing the data between 2002 and 2018.} %As presented in Fig. \ref{fig:X7_2016_NACO} and Fig. \ref{fig:distribution_gasdust_s50}, the intensity maximum of the dust is located at the position of S50 in the data \textbf{until} 2008. The gas, as shown partially in Fig. \ref{fig:turnover_point} and more detailed in Fig. \ref{fig:distribution_gasdust_s50} is symmetrically arranged around S50, coinciding with the intensity maximum of the dust. The intensity maximum of the gas can clearly be observed in the head of X7 (Fig. \ref{fig:turnover_point}). Followed by the observation of X7 and S50 in 2018, we find that the bright $L'$-band dust and Br$\gamma$ emission show a prominent asymmetry. While the bright Br$\gamma$ emission is found in the head of X7, the bright dust emission is located behind the head and therefore closer to S50 (Fig. \ref{fig:distribution_gasdust_s50}).

\subsection{Photometric analysis of X7}
\label{sec:mag}
The photometry was done in the H-, K-, $L'$-, and $M'$-band.
As shown in Fig. \ref{fig:X7_2016_NACO}, Fig. \ref{fig:distribution_gasdust_s50}, Fig. \ref{fig:comic_1999}, and Fig. \ref{fig:front_tip_distance} the dusty \textbf{bow shock} of X7 gets elongated between 1999 and 2018. After 2007, the projected elongated size of the $L'$-band emission exceeds a spatial coverage of two PSF ($\approx\,0.20"$) and we categorize the source in a front- (i.e. head) and back-part (i.e. tail). For this analysis, we focus on the tail of X7 since deriving the emission area of the faint $L'$-band head magnitude is not free of confusion. For the photometric analysis, we use S65 because of its \textbf{well-known} stable magnitude of about $10.96$ mag \citep{Hosseini2020}. For the magnitude of X7, we use the peak emission of the $L'$-band dust emission (see Fig. \ref{fig:X7_2016_NACO}). The magnitude of X7 is derived \textbf{from} the peak intensity and can be related to the tail of the source after 2008. For every \textbf{dataset}, a one-pixel aperture is used. No background subtraction is applied because of the high S/N ratio that exceeds several orders of \textbf{magnitude} the intensity of the surroundings.\newline 
The fit presented in Fig. \ref{fig:variable_X7_mag} can be categorized in two different results:
\begin{figure}[htbp!]
	\centering
	\includegraphics[width=.5\textwidth]{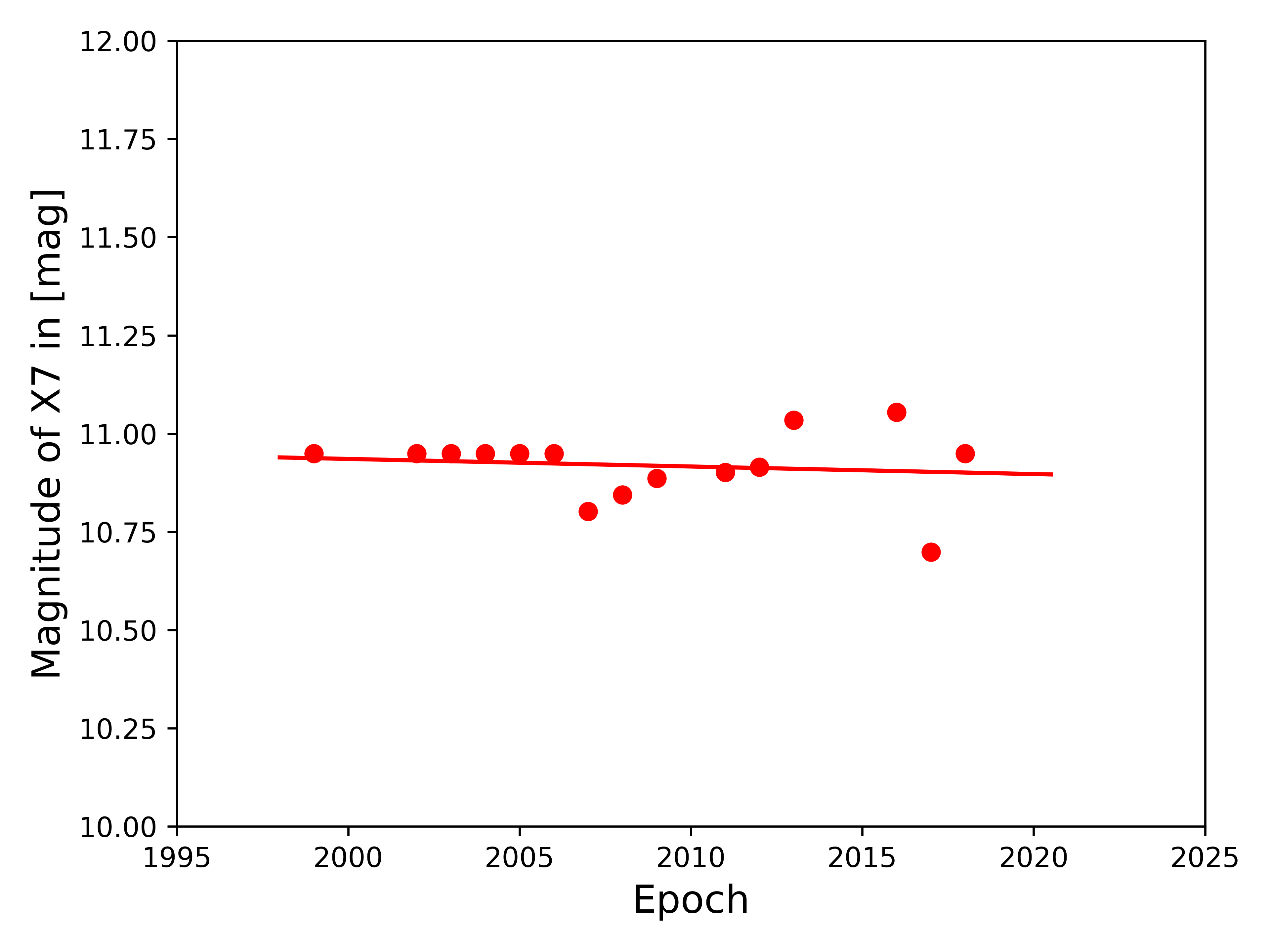}
	\caption{$L'$ magnitude of X7 between 1999 and 2018 with a typical uncertainty of $\pm\,0.02$ mag \citep[see also][]{Hosseini2020}. The data before 2002 was observed with COMIC/ADONIS+RASOIR and presented partially in \cite{clenet2001}. The red data-points shows the magnitude of X7 till 2007. After 2007, the main peak emission can be found in the back of the emission source and is therefore related to the tail of X7. In this figure, a lower magnitude value is brighter.}
\label{fig:variable_X7_mag}
\end{figure}
\begin{enumerate}
    \item A constant magnitude of X7 before 2007,
    \item A variable magnitude of the tail after \textbf{2007}.
\end{enumerate}
Regarding point 1, the COMIC/ADONIS+RASOIR and NACO $L'$-band data between 1999 and \textbf{2006} does not show a magnitude variation. Additionally, point 2 underlines a slightly variable $L'$-band \textbf{tail magnitude} of the \textbf{bow shock} at the K-band position of S50.\newline
These variations of the $L'$-band magnitude of X7 coincide with the discontinuous shape evolution that is observed in the Br$\gamma$ \textbf{line maps} (see Fig. \ref{fig:turnover_point} and Fig. \ref{fig:distribution_gasdust_s50}).\newline
By investigating several \textbf{datasets} of the GC that cover individual bands, we 
\begin{table}[htb!]
%\caption{}
\tabcolsep=0.1cm
\begin{center}
\begin{tabular}{ccccc}\hline \hline
Band & Central wavelength & mag$_{S2}$&mag$_{S50/X7}$ & flux$_{S50/X7}$ \\
     & in [$\mu m$] & &  & in [mJy]\\  \hline
H  & 1.65 & 16.00 & 19.65   & 0.0861 \\
K  & 2.20 & 14.13 & 16.00   & 0.2459 \\
$L'$ & 3.80 & 11.33 & 10.72   & 12.680 \\  
$M'$ & 4.80 & 12.3  &   9.12 & 33.736 \\
\hline
\end{tabular}
\end{center}
\caption{Magnitude and flux of the \textbf{bow shock} source X7. We use the band related ESO filter for the zero magnitude flux. The dereddened H-, K-, and $L'$-band values are related to the SINFONI and NACO data of 2016. The $M'$ data-point is determined from the NACO data of 2012 where we applied a flux conserving smooth-subtract Gaussian (PSF-sized kernel).}
\label{tab:mag_S2_X7_S50}
\end{table}
find an increasing flux towards higher bands (from H- to M-band, see Table \ref{tab:mag_S2_X7_S50}) for X7. Using the magnitude values, we derive the SED with a two-component fit for the emission of S50 (H and K) and X7 ($L'$ and $M'$).
\begin{figure}[h!]
	\centering
	\includegraphics[width=.45\textwidth]{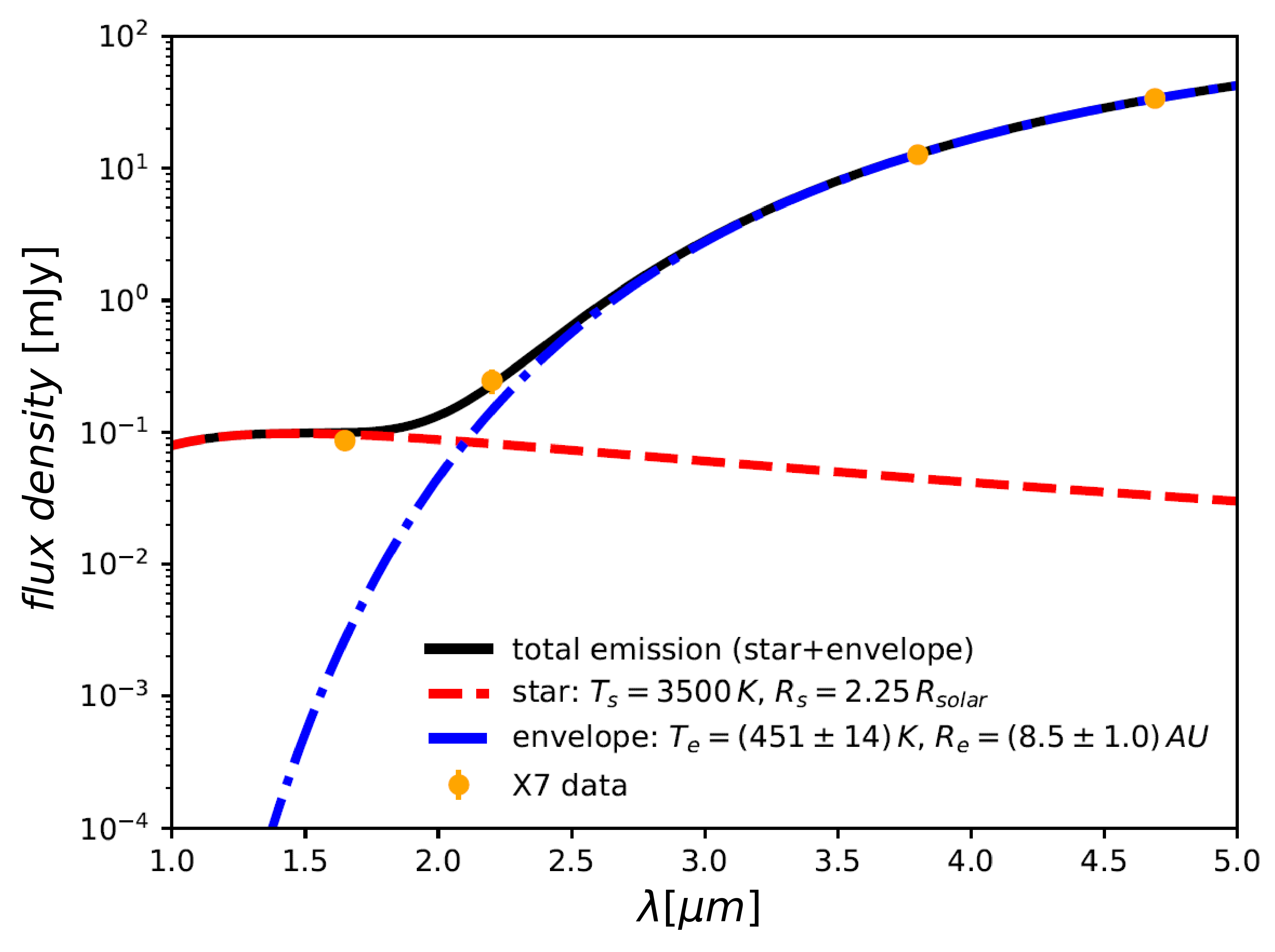}
	\caption{Spectral energy distribution of the X7/S50 system that indicates a dust-embedded stellar source.}
\label{fig:SED}
\end{figure}
This indicates a dust-dominated emission source with a \textbf{multiwavelength} appearance. \textbf{Since the commonly observed dust temperature in the GC is about $200$ K \citep{Cotera1999}, the derived envelope temperature of $~450$ K must be heated up by the internal stellar source S50.}

\section{Discussion and Conclusion}
\label{sec:DiscussionConclussion}
In this section, we will discuss the results and the implications for future observations of the X7/S50 system. We will also speculate about some possible interpretations \textbf{regarding} the \textbf{increasing} position angle and the \textbf{implied decoupling of X7 and S50.}

\subsection{The shape of X7}

From the survey of X7 over two decades \textbf{with all publicly} available SINFONI and NACO data, we \textbf{have shown} that the shape of the \textbf{bow shock} does change over time on a significant level. Even when we consider different weather and background scenarios, the \textbf{here presented} findings \textbf{underline} a dynamical star-envelope setup. As shown in Table \ref{tab:X7_length} and Fig. \ref{fig:turnover_point}, the shape of X7 undergoes a transition: we find an almost constant position angle and magnitude with a linear increasing \textbf{bow shock} size both in gas and dust till 2009. Based on \cite{muzic2010}, this setup for X7 is expected because S50 as the stellar counterpart is located close to the front tip of the \textbf{bow shock} X7. As theoretically described by \cite{Wilkin1996, Wilkin2000} and observed by \cite{muzic2010}, we can confirm that the S-star S50 is located always at the position of the maximum peak intensity of the observed $L'$-band emission of the \textbf{bow shock} X7. This $L'$-band intensity peak can be found close to the \textbf{apex} of the \textbf{bow shock} at a distance of $R_0\,=\,\sqrt{\frac{\dot{m}_{\rm w} v_{\rm w}}{\Omega \rho_{\rm a} v^2_{\rm a}}}\,\approx\,2.5\,\times\,10^{15}$cm \citep{muzic2010} till 2009. Here, $\dot{m}_{\rm w}$ describes the mass-loss rate of the star, $v_{\rm w}$ is the stellar wind velocity, $\Omega$ a dimensionless parameter to control the shape of the \textbf{bow shock} ($\Omega=4\pi$ for an isotropic stellar wind), $\rho_{\rm a}$ is the density of the ambient medium, and $v_{\rm a}$ the relative stellar velocity in a non-stationary medium. 

Between 2009.47 and 2010.49, we observe a discontinuous process since the \textbf{Br$\gamma$ and $L'$-band} size is decreased by almost $30\%$ compared to the observation in 2009.26 (NACO). After 2010, not only is the Br$\gamma$ and $L'$-band continuum size expanding, but also the position of S50 seems to change with respect to the shell. Hence, $R_0$ is not a fixed value anymore and seems to change year by year. Because the stellar position \textbf{with respect to its dusty envelope} does not follow any simple stationary model, we will speculate about some possible interpretations.

The authors of \cite{Henney2019} discuss dust- and bow-waves as a possibility for asymmetric shapes. Considering a possible `rip-point' (where the shell gets detached from the star) harbors the \textbf{problem that} these processes (including the trajectories of the \textbf{dust grains}) take up to several 1000 years as proposed by \cite{Henney2019}. We have shown that the gas distribution is coinciding with the dust emission (see Table \ref{tab:X7_length} and Fig. \ref{fig:distribution_gasdust_s50}). In 2008, we find a matching size of the emission of about 230 mas. The NACO data of 2009.26 seems to follow the linear evolution of the observed emission size in 2008. For the SINFONI data of \textbf{2009.47,} we observe a source size that \textbf{is unexpected}. Because of these timescales, we see a reduced chance for the possibility of dust- and \textbf{bow-waves} as a suitable explanation for the discontinuous evolution.

Another possibility are projection effects. Considering the \textbf{possibility that} S50 could maybe not be related to X7 at all and just moves on a random orbit that coincides in projection with X7 opens a new set of questions. In the \textbf{following,} we independently discuss these questions ignoring the already complete discussion of \cite{muzic2010}, where the authors exclude the possibility of a random encounter based on the matching proper motion of S50 and X7.

The most obvious one is regarding the statistical robustness of a randomized orbit that is \textbf{oriented} along the trajectory of X7 over \textbf{time.} As derived by \cite{Sabha2012} and \cite{EckartAA2013}, the probability for such an event is of the order of $10^{-4}$ to a few percent for a consecutive observation of 3 years. The probability for the outer \textbf{region of the S-cluster} should therefore be in a comparable range since we observe S50 along with X7 between 2002-2009 (NACO) and 1999 with COMIC/ADONIS+RASOIR (Appendix, Fig. \ref{fig:comic_1999}). 

\textbf{As shown in Fig. \ref{fig:X7_2016_NACO} and Fig. \ref{fig:distribution_gasdust_s50}, the shifted $L'$-band intensity maximum towards the tail is followed by the projected position of S50. Based on the derived $L'$-band magnitude year-by-year, the temperature of X7 is always well above $200$K which can only be achieved by an internal heating source. Hence, we conclude that the tail of X7 gets heated up by S50. Alternatively,} a wind that originates \textbf{south-west of the position of Sgr~A* could be responsible for the increased tail emission in 2018}. However, this does not explain the Wilkinoide \citep{Wilkin1996} \textbf{bow shock} between 2002 and 2009 that is observed throughout the NACO and SINFONI data. In combination with the continuum and \textbf{line emission} data of 2006 and 2008 (see Fig. \ref{fig:distribution_gasdust_s50}), we will not discuss the possibility of another wind coming from south-east any further, \textbf{especially} considering the observed footprint \textbf{of a wind that originates at the position of Sgr~A* or IRS16} in the mini-cavity \citep[see, e.g.,][]{Lutz1993}.\newline

A more suitable explanation of the observed gas and dust emission \textbf{of X7} is forward scattering explained by the Mie-theory. This scatter mechanism describes \textbf{dust grains as an emitter} with the mentioned forwarded scattering. \textbf{Single and multiscattering events occur} where, e.g., dust emits and transmits stellar \textbf{light, which is reemitted} by close-by grains. As long as S50 is embedded in the dusty shell X7, the ionized and blue-shifted Br$\gamma$-emission is symmetrically distributed following the aligned dust grains. After 2009, the peak emission of the $L'$-band emission can be observed closer to the tail of X7 whereas the gaseous tip gets more prominent\footnote{We advise the interested reader to compare the Br$\gamma$ emission of 2008 and 2018 presented in  Fig. \ref{fig:distribution_gasdust_s50}.}.  

Overall, we conclude that a projection scenario that describes a random encounter between S50 and X7 is highly unlikely but not excluded.

\subsection{Two observed processes: the change of the position angle between X7 and Sgr~A*}
\label{sec:section_pa}
Besides the observed decreased projected source size in 2009-2010, we find that the position angle \textbf{(with respect to Sgr~A*)} is increasing faster as the shell of S50 is \textbf{aligned} towards the SMBH (Table \ref{tab:X7_length}). Even though a change of the position angle is expected since the proper motion of the X7/S50-system is directed towards \textbf{the} north \citep{muzic2010}, \textbf{the gas and dust shell} is pointing/aligned to a position $0.45"$ north of the SMBH (see Fig. \ref{fig:X7_2016_NACO}, \ref{fig:distribution_gasdust_s50}) in 2018. Comparing the position angle of 2006 and 2018 shows a growth \textbf{of} about $40\%$. If S50 would be located close to the position of the tip of the \textbf{bow shock} at a distance $R_0$, a growth by around $12\%$ would be expected in 2018. However, assuming the chance of reading uncertainties, the position angle of $60^{\circ}$ in 2018 between Sgr~A* and X7 marks a lower limit. The observations and the measured properties \textbf{suggest to} distinguish the description of X7 in pre 2009.26 and post 2009.46 since the object shows a discontinuous development as a function of time.\newline Summing up the observational results leads to two assumptions: either X7 is a tidally stretched object \footnote{Discussed by Randy Campbell et al., UCLA, at GCWS 2019 (proceedings in prep.).} where the head is on its way towards Sgr~A* (A), or the dust- and gas-shell seems to be ripped apart by an unknown interaction (B).\newline\newline 
\begin{itemize}
    \item[A)] 
The trajectory of the head, as shown in Fig. \ref{fig:front_tip_distance}, shows a clear trend towards Sgr~A*. The distance between the SMBH and the gaseous head of X7 decreased by around $20\%$ over almost two decades. Taking into account the proper motion of the S50/X7 system, this is expected. Even though a clear trend can be observed, projection effects could also play a role because of the orbit of S50 (see Appendix, Fig. \ref{fig:comic_1999}). Studying the projected positions of the head, tail, and S50 with respect to Sgr~A* (Fig. \ref{fig:front_tip_distance}) implies that the R.A. distance of the head stays almost constant. If the head would be attracted by Sgr~A*, we would not observe a preserved dusty shell of X7 because the front would simply accelerate towards the SMBH with respect to S50 and the tail. Hence, the shape of the Br$\gamma$-emission in 2018 might be explained by the forward (and backward) single- and \textbf{multiscattered} stellar light of S50. If upcoming observations can confirm the observed decoupling of the head from S50 and its tail, it might trigger the flaring activity of Sgr~A* above the statistical level \citep{Witzel2012}. Please consider also the Appendix (Fig. \ref{fig:X7_strechted}) for a possible outlook.
\end{itemize}
\begin{itemize}
    \item[B)] 
As discussed before, the Br$\gamma$ \textbf{line map} of 2009 (Fig. \ref{fig:turnover_point}) but also the size of the $L'$-band continuum detection (Table \ref{tab:X7_length} and Fig. \ref{fig:X7_2016_NACO}) marks a noticeable step in the discontinuous evolution of X7. Adding the growth of the position angle of X7, the increasing distance between the head and S50, and the relative position of the shell and the S-star to the calculation creates the assumption that we observe a dissolving event. Since the overall shape of the dust shell as observed with NACO seems to be preserved even though a \textbf{clearly} increased elongation can be observed, it is safe to assume that \textbf{the shell} stays intact. Hence, clear evidence for the scenario of a destroyed shell cannot be given.
\end{itemize}

Considering the here discussed observational results leads to the problem of the ongoing spatial misplacement of S50 with respect to X7 and the growing position angle. We will elaborate on this in the following subsections.

\subsection{Unexpected event around 2010}

Recently, \cite{Vorobyov2020} modeled the behavior of gas and dust features of protoplanetary disks which move with a supersonic motion in a dense ambient medium. Considering the Br$\gamma$ emission in Fig. \ref{fig:turnover_point} in 2009 in combination with the related $L'$-band emission size (Table \ref{tab:X7_length}), we conclude that there might be a prominent decoupling of gas and dust as discussed by \cite{Henney2019}. As \textbf{discussed}, the time scales of the cited work does not fit the observation. Hence, the observations suggest the presence of a disturbing event. We speculate \textbf{that this} event \textbf{has} been caused by the close fly-by (in projection) of \textbf{S33,} which would at least partially explain the \textbf{almost compact} Br$\gamma$ line map emission in 2009 and the discontinuous evolution of the projected $L'$-band size of X7 (see Table \ref{tab:X7_length}). A critical parameter of this speculative scenario is the 3-dimensional distance and therefore \textbf{the position} of S50/X7 and S33 with respect to each other. 

For giving an estimate on the 3-dimensional distance between S33 and S50, we use the related proper motion ($v_{t}$) and line-of-sight velocity ($v_{r}$). For $v_{r}$, we use a lower limit of around 500 km/s \citep{muzic2010}. For deriving a LOS velocity, an averaged value of the observed H$_2$Q(1)\footnote{Transition v=1-0 Q(1)} and H$_2$Q(3)\footnote{Transition v=1-0 Q(3)} absorption line is used. Hence, for $v_{t}$ of S50 we derive a value of around 350 km/s in 2018 (see Appendix, Fig. \ref{fig:H2_emission} and Table \ref{tab:emissionlines}). This velocity \textbf{estimate results} in a approximate 3-dimensional velocity of $({v_{r}}^2+{v_{t}}^2)^{-1/2}\approx\,600 \, km/s$. This results in an approximate distance $d$ towards Sgr~A* of $d_{S50}\,\approx\,0.047\,pc\,\approx\,1.19"$. From \cite{Ali2020}, we use the 3-dimensional position of S33 based on their presented orbit plots. We find that the 3-dimensional distance of S33 in 2009 with respect to Sgr~A* is about $1.2"$. Because the 3-dimensional distance of S50 with respect to Sgr~A* is a lower limit, we set the distance of S33 to S50 at about $0.01"$ or $120\,AU$.\newline
Considering the derived 3-dimensional distance between S33 and S50, the modeled interaction between an intruder and the \textbf{host star} with an envelope as presented in \cite{Vorobyov2020} could be a possibility. A detailed model should answer the question about the stellar-wind interaction with the ambient wind \citep{Yusef-Zadeh2020} but exceeds the scope of this work.

Furthermore, it should be mentioned that \cite{Gorman2015} and \cite{Wallstrom2017} presented ALMA observations which do not show a symmetrical dust/gas distribution of the envelope related to the \textbf{host star} (which happens to be in both cases a giant). \cite{Wallstrom2017} observed a \textbf{so-called} 'Spur' which describes an asymmetric \textbf{gas feature} related to the host star. This 'Spur' could be compared to the dust and gas shell X7 of S50. \cite{Wallstrom2017} \textbf{argue} that this 'Spur' might be created by a sporadic eruption event of the host star. Nevertheless, \cite{Zajacek2020} modeled recently the depletion of \textbf{red giants} and showed that the detached and shocked envelope of the \textbf{host star} can suffer from the interaction with Sgr~A*. Even though \cite{Schartmann2018} used stellar winds to model the S2 peri-center passage, it is shown that the presence of a SMBH results in an asymmetric mass distribution. If the gas/dust shell got detached and its length scale increased beyond the stellar Hill radius, the gravitational influence of Sgr~A* would dominate the evolution of X7 as was described by \cite{Eckart2013a} and numerically modelled by \citet{Zajacek2014}. 

\subsection{The nature of the source X7/S50}

From the \textbf{multiwavelength} analysis with NACO and SINFONI in the H-,K-,L, M-band, and the modeled \textbf{SED,} we find that the X7/S50 system consists of a stellar component in combination with the internally heated dusty envelope (Fig. \ref{fig:SED}).\newline
Comparing the SINFONI Br$\gamma$-line map of 2006 and 2018 with the NACO $L'$-band continuum observations shows the gas- to dust-component ratio is around 1:2-1:3 which are typical values for HAe/Be or T-Tauri stars \citep{Mannings2000}. The weak H$_2$-absorption lines (Appendix, Fig. \ref{fig:H2_emission}) underline the possibility for observing a YSO as discussed in \cite{muzic2010}. The theoretical modeling of the dust and gas of X7 strengthen the possibility of a YSO.\newline
Additionally, \cite{Rivinius1997} \textbf{reported wind variations} for early-B hypergiants with \textbf{mass-loss} rates of several $10^{-6}\,{\rm M_{\odot} yr^{-1}}$. This variations are also investigated by \cite{Muratorio2002}. In both cases, the P-Cygni profile of highly excited $[{\rm FeIII}]$ multiplets/lines are indicators for a complex wind interaction with the stellar source. Even if we do not find a prominent P-Cygni profile in the spectrum, a \textbf{nondetection} can be explained by the high sky emission line variations which leads to over/undersubraction effects as shown by \cite{Davies2007}. Finding a P-Cygni feature would increase the complexity of the X7 system since there would be wind-wind-accretion processes that should be a part of \textbf{the mentioned} model. The wind launched at the position of Sgr~A* would be accompanied by stellar winds of S50. Therefore, the S50 dust and gas accretion would be influenced by the aforesaid wind-wind process.

Furthermore, the origin of the excited $[{\rm FeIII}]$ lines is still not clear \textbf{\citep{Peissker2019, Peissker2020c}} even though we speculate the detection could be linked to the area and the Br$\gamma$-bar \textbf{\citep{Schoedel2011, Peissker2020d}}. However, \cite{Wolf1985} mention that higher excited $[{\rm FeIII}]$ lines could have been pumped by HeI lines. In the spectrum of X7, we find a strong \textbf{blue-shifted} HeI line at $2.058\,\mu m$\footnote{Transition $2p^1 P^0-2s^1S$} with a matching LOS velocity. Hence, we consider the pumping of the forbidden Fe-lines as a possible explanation. For the sake of completeness, we note that every of the four most prominent emission lines in the present K-band spectrum in Fig. \ref{fig:X7_overview_2018} is accompanied by a less intense line which is related to the source X7.1/G5. \textbf{In addition,} we do find a red-shifted H$_2$ line (about $650\,km/s$) at $2.228\,\mu m$ (transition v=1-0 S(0)). Because of the direction of the Doppler-shifted H$_2$ line, this emission might probably be related to another species.\newline
From the here shown results and \textbf{the discussed} scenarios, we conclude \textbf{that the} stellar source of X7 can be associated without any doubt \textbf{with} the S-cluster star \textbf{S50,} which \textbf{confirms} the analysis of \cite{muzic2010}. As implied by the H$_2$ absorption lines, the LOS velocity of the star is blue-shifted. Hence, the Doppler-shifted direction of the stellar LOS-velocity matches the \textbf{emission lines} of the surrounding envelope, which also shows a blue-shifted motion.\newline 
The shape of the \textbf{bow shock} in 2002 is almost spherical and Wilkinoide. With the presented COMIC/ADONIS+RASOIR data of 1999, we find evidence that earlier $L'$-band data than 2002 confirm the trend of a `growing' dusty envelope.\newline 
The two distinct observed processes, the LOS-velocity, and the star/envelope evolution \textbf{underline} the prominent dynamical process that highlights the \textbf{uniqueness of the} X7/S50 system.\newline 

Along the X7/S50 \textbf{source,} we observe a strong and prominent velocity gradient in 2018. Considering the existence of a formed wind at the position of Sgr~A* or IRS16, we assume this might be the origin of the \textbf{gradient. In} 2009, it seems the envelope starts to interact with the nearby S-cluster star S33 since we trace indications of this possible interaction in the same year (Fig. \ref{fig:turnover_point}). 
The $L'$ NACO data \textbf{shows that} the tail of X7 gets brighter between 2010 and 2018. We predict that this gain of brightness will likely continue in the future. 
We also \textbf{speculate that} the ongoing interaction \textbf{of} S33 and Sgr~A* with the shell of S50 could lead to the partial destruction of the \textbf{bow shock}.

% An external wind could be responsible for a velocity gradient that is observed along the source from \textbf{the tip} to tail.

\subsection{Sporadic or stellar winds?}

As we have observed and presented in Fig. \ref{fig:X7_2016_NACO} but also listed in Table \ref{tab:X7_length}, the shell of S50 is pointing in projection above Sgr~A*. As proposed by \cite{Wardle1992}, strong stellar winds arising from the IRS16 complex are responsible for the creation of the mini-cavity. The authors discuss an observed $2.217 \mu m$ emission line at the position of the mini-cavity \citep[see also][]{Lutz1993} which can most likely be related to the [FeIII] multiplet observed in several dusty sources west of Sgr~A* \textbf{\citep{Ciurlo2020, Peissker2020c}}. 
\begin{figure}[htbp!]
	\centering
	\includegraphics[width=.5\textwidth]{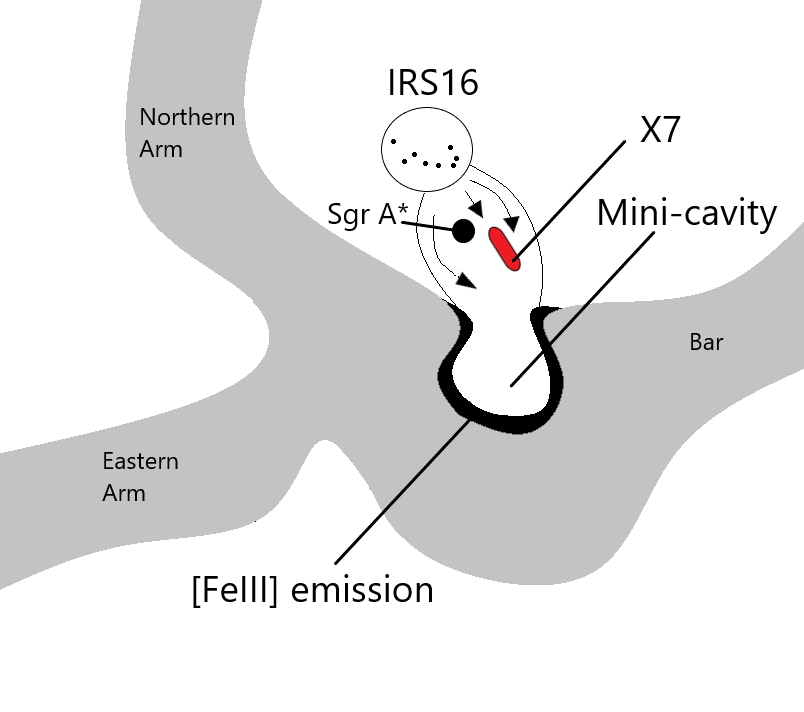}
	\caption{Sketch adapted from \cite{Wardle1992}. The [FeIII] emission is also observed by \cite{Lutz1993}. The position of X7 and the observed position angle of 2018 is implied with the red object. Sgr~A* is located at the black dot.}
\label{fig:irs_wind}
\end{figure}
The ionized iron multiplet can also be observed for X7/S50 as shown in Fig. \ref{fig:X7_overview_2018}. If we exclude the possibility of a wind arising at the position of Sgr~A*, the excitation of iron as well as the position angle (Table \ref{tab:X7_length}) could be linked to stellar winds from IRS16. The supermassive black hole would be responsible for \textbf{refocusing} the wind (Fig. \ref{fig:irs_wind}) and \textbf{sources that} are leaving the `slip-stream of Sgr~A*' would suffer from this interaction.
This dynamical evolution of the gaseous and dusty shell of the X7/S50 system underlines the need for a constant survey of the GC region in various bands.\newline\newline
If a wind responsible for the alignment and evolution of the X7/S50 system is indeed arising at the position of Sgr~A*, the apparent change of the position angle with respect to Sgr~A* is unexpected. Since we clearly observe the evolution of the elongation of the X7/S50 system, it may be explained by a temporarily active wind phase of Sgr~A* as indicated by \cite{morris1996}. Speculatively, this could contribute to the `Paradox of youth' \citep{Ghez2003} where star formation is `allowed' for a short period of time. Nevertheless, in combination with the X7 proper motion \citep{muzic2010} directed towards \textbf{the north,} the alignment angle of X7/S50 may have been induced to the system before 2009. After 2009, the wind activity may have been decreased while the position angle increased (Table \ref{tab:X7_length}) because of the proper motion of the X7/S50 system.

\subsection{Future observations with the Extremely Large Telescope and the James Webb Space Telescope}

Near- and mid-infrared instruments will play a key role in investigating the evolution of the X7/S50 system. The prominent detection of X7 in the $L'$- and $M'$-band promises successful observations with MIRI \citep[James Webb Space Telescope, see][]{Bouchet_2015, Rieke_2015, Ressler_2015}, METIS \citep[Extremely Large Telescope, see][]{Brandl2018}, and MICADO \citep[Extremely Large Telescope, see][]{Trippe2010}. MIRI and METIS will be able to finalize the investigation about the possible clumpiness of X7 which could be used for theoretical models \textbf{\citep[e.g., the filling factor, see][]{Peissker2020d}}. With a more accurate result, we will be able to precisely determine the density and therefore the mass of the dusty shell. Furthermore, we are able to search for more complex emission lines in the local line of sight ISM \textbf{like, for example,} $NH_3$. \textbf{Additionally,} gas emission lines like, e.g., $CO$ and $HCN$, can provide a more detailed description about the nature of the X7/S50 system. These gas- and ice-absorption lines can also be used as an additional probe for a stellar disk and a possible YSO. \cite{Moultaka2006} and \cite{Moultaka2009} showed that these lines are useful to determine local extinction values for the interstellar medium \textbf{\citep[see also][]{Schoedel2010, Peissker2020d}}.\newline Even if we have shown S50 can be associated with the stellar counterpart of X7, a hidden star at \textbf{a distance} of $R_0$ from the apex of the \textbf{bow shock} should be detectable with MICADO \citep[see the simulated view of the GC with MICADO in][]{Davies2010}.\newline
As we have presented in Fig. \ref{fig:X7_2016_NACO}, investigating the GC with a wider FOV in the mentioned bands should also reveal more (elongated) sources that might be suffering from the wind that is formed at the position of Sgr~A* or at IRS16. 
We conclude that the upcoming observations of the GC with the ELT will be able to manifest the dynamical influence of the nuclear wind. We can safely assume the X7/S50 system will not be the only source in the GC which is undergoing a dynamical influence. \cite{Yusef-Zadeh2017} already showed that YSOs with bipolar \textbf{outflows} can be observed in the environment of the SMBH. Even though we cannot finally answer the question about the nature of the X7/S50 system, we see some weak traces that point towards its YSO nature.
If \textbf{the theoretical} models reveal matching parameters of the X7/S50 system with a YSO, the origin of these sources is still not clear. \textbf{However,} the implication of a population of YSOs promises an important cornerstone in the investigation of the direct vicinity of the nearest SMBH that resides in our Galaxy.

\acknowledgements%We highly appreciate the comments of the anonymous referee that helped to improve this paper.
This work was supported in part by the
Deutsche Forschungsgemeinschaft (DFG) via the Cologne
Bonn Graduate School (BCGS), the Max Planck Society
through the International Max Planck Research School
(IMPRS) for Astronomy and Astrophysics as well as special
funds through the University of Cologne. Conditions and Impact of Star Formation is carried out within the Collaborative Research Centre 956, sub-project [A01], funded by the Deutsche Forschungsgemeinschaft (DFG) – project ID 184018867. FP is grateful for the child care support of the DFG. MZ acknowledges the financial support by the National Science Center, Poland, grant No. 2017/26/A/ST9/00756 (Maestro 9) and the NAWA financial support under the agreement PPN/WYM/2019/1/00064 to perform a three-month exchange stay at the Charles University in Prague and the Astronomical Institute of the Czech Academy of Sciences. Part of this
work was supported by fruitful discussions with members of
the European Union funded COST Action MP0905: Black
Holes in a Violent Universe and the Czech Science Foundation
-- DFG collaboration (No.\ 19-01137J). JC, SE, and GB contributed useful points to the discussion. We also would like to 
thank the members of the SINFONI/NACO/VISIR and ESO's Paranal/Chile team for their support and collaboration.

\bibliographystyle{aasjournal}
\bibliography{bib.bib}

\appendix
\label{sec:appendix}
\section{K-band position of S50 in relation to the $L'$-band emission of X7}

\begin{figure*}[htbp!]
	\centering
	\includegraphics[width=1.\textwidth]{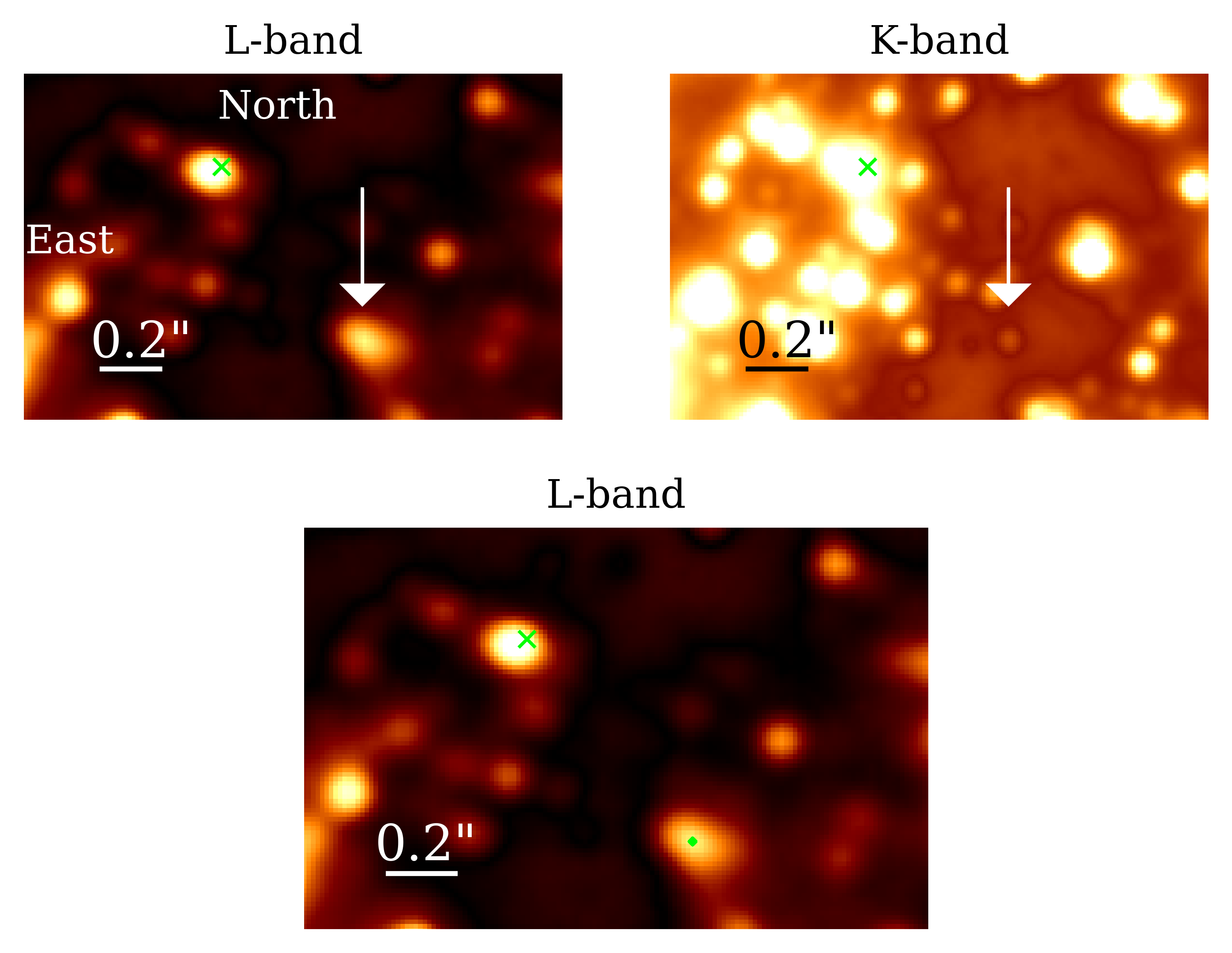}
	\caption{Galactic center observed with NACO in the L- and K-band in 2002. \textbf{In the upper left and right panel,} Sgr~A* is indicated with a green $\times$, the white arrow points towards the position of X7 (L-band) and S50 (K-band). As in Fig. \ref{fig:comic_1999}, S65 can be used as a reference source for the identification. \textbf{With the combination of the $L'$- and K-band data, we derive the position of the stellar source S50 with respect to X7 (lower panel, see the green dot inside the dusty emission). For the interested reader, we note that the} K-band image also demonstrates the high asymmetrical stellar distribution of the S-cluster in projection.}
\label{fig:KL_ident_naco_2002}
\end{figure*}

Here we are showing the relation between the K-band detection of S50 and the $L'$-band emission of X7 (Fig. \ref{fig:KL_ident_naco_2002}) observed with NACO. To compare the projected \textbf{on-sky} distances, we are rebinning the $L'$-band data to the same pixel scale as the K-band data, i.e., two pixels correspond to 27 mas.
By using the stellar position of S50 in the K-band, we pinpoint the stellar location in the $L'$-band (see Fig. \ref{fig:KL_ident_naco_2002}). This procedure is similar to the steps for the SINFONI detection with the difference that we are using \textbf{data cube}s. In the final mosaic data cube of a related year, we select the $2.0\,-\,2.2\,\mu m$ range to extract the related K-band image. Then, we compare the position of S50 in the extracted K-band image with the continuum subtracted Br$\gamma$ line maps that are constructed from the related data cube (Fig. \ref{fig:distribution_gasdust_s50}).

\section{$H_2$ emission of S50}
\label{appendix_h2_emission}
For investigating the spectrum of S50, two main cornerstones have to be fulfilled:
\begin{enumerate}
    \item A maximized data quality,
    \item An individual detection of S50.
\end{enumerate}
Regarding point 1, a high number ($>\,20$) of single exposures with a satisfying quality (FWHM $<\,6.5$ pixel in x- and y-direction) results in an increased S/N ratio. Using the SINFONI data in 2018 (Table \ref{tab:data_sinfo5}) fulfills this first requirement. The second point is limited by nature. Using data where S50 coincides with its shell could lead to \textbf{a confused and blended} spectrum. However, studying the projected position of the stellar counterpart of the dusty and gaseous shell X7 \textbf{reveals} the data in 2018 \textbf{matching} the needed conditions (see Fig. \ref{fig:distribution_gasdust_s50}).
\begin{figure*}[htbp!]
	\centering
	\includegraphics[width=.5\textwidth]{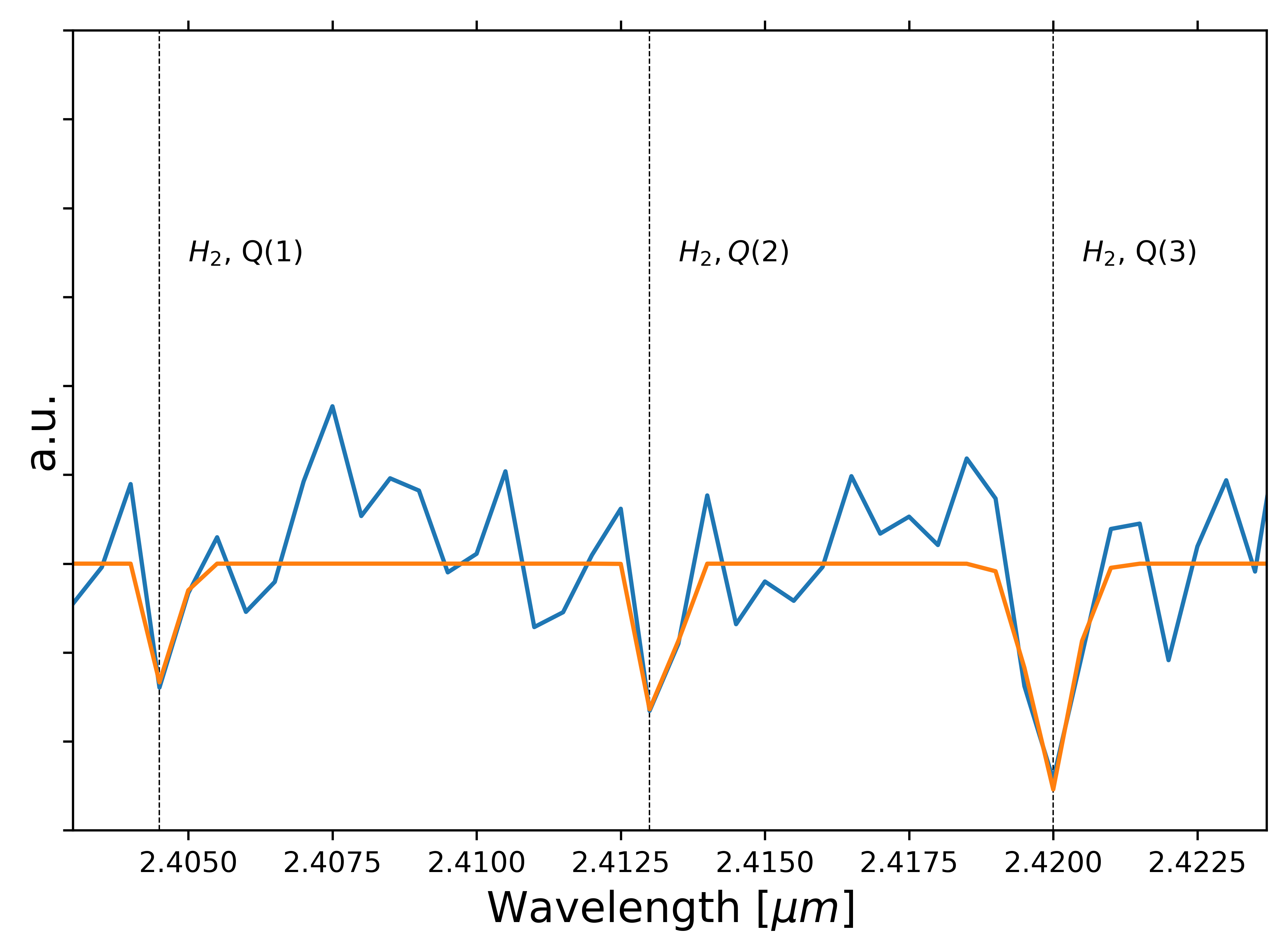}
	\caption{H$_2$ triplet measured at the K-band position of S50 with SINFONI.}
\label{fig:H2_emission}
\end{figure*}
For the spectrum presented in Fig. \ref{fig:H2_emission}, we use a PSF sized aperture. Furthermore, we fit a Gaussian to the detected H$_2$ triplet with a measured uncertainty of about $\pm\,35$ km/s. As pointed out by, e.g., \cite{Arulanantham2017} and \cite{Hoadley2017}, H$_2$ lines can be used as a tracer for protoplanetary disks of YSOs. Considering the analysis of \cite{muzic2010} and the proposed nature of S50 as a T-Tauri or Ae/Be Herbig star seems to be a reasonable connection. However, we would like to point out that future observations in combination with theoretical models will confirm or reject this claim.

\section{X7, a tidally stretched feature}
As a rather speculative scenario, we shortly discuss the possibility that X7 is a tidally stretched gas and dust feature (as proposed by Randy Campbell, UCLA, at GCWS 2019; proceedings in prep.). Isolating the observation of the X7/S50 observation in 2018 could lead indeed to the assumption that the source is a tidally stretched gas and dust feature. Even though this scenario promises a wide range of useful scientific implications, observations of comparable objects have shown that a tidally stretched object is rather unlikely \citep{Gillessen2012, EckartAA2013, Valencia-S.2015}.
\begin{figure*}[htbp!]
	\centering
	\includegraphics[width=.5\textwidth]{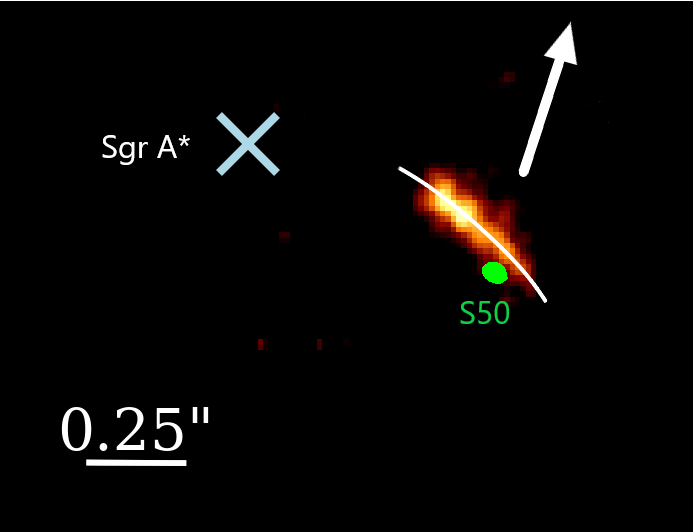}
	\caption{Possible evolution of the X7/S50 system. The sketch is based on the \textbf{line map} detection of X7 in 2018 at about $2.161\,\mu m$ shown in Fig. \ref{fig:distribution_gasdust_s50}. The green filled contour represents the K-band position of S50. The white curved line along X7 corresponds to the speculative scenario where the head of the system gets detached and attracted by Sgr~A* (light-blue x). The arrow indicates the direction of the proper motion of X7 and S50 as derived by \cite{Gillessen2009} and \cite{muzic2010}.}
\label{fig:X7_strechted}
\end{figure*}
Considering Fig. \ref{fig:front_tip_distance} (left side), we do find an increasing distance of the head from S50. However, the overall trend of the X7/S50 system seems to be not affected by Sgr~A* (Fig. \ref{fig:front_tip_distance}, right side). Even with the observed and detected asymmetry regarding the stellar position with respect to its gaseous and dusty shell X7, the system is following the proper motion as found by \cite{muzic2010}. As pointed out several times, a \textbf{long-time} survey of the evolution of X7/S50 is required.

\section{COMIC/ADONIS+RASOIR data of 1999}
\label{subsec_COMIC_ADONIS}
\begin{figure*}[htbp!]
	\centering
	\includegraphics[width=1.\textwidth]{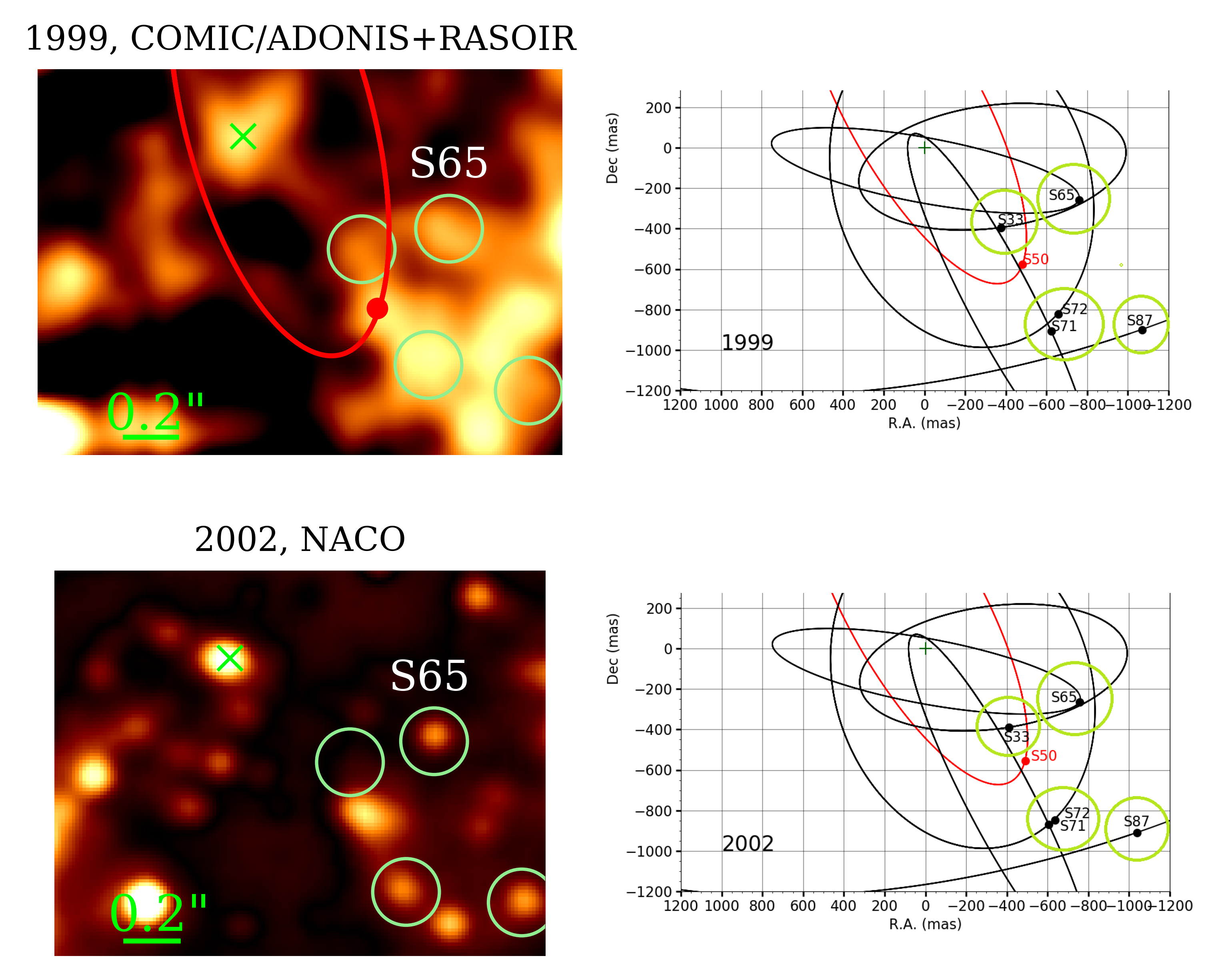}
	\caption{Galactic center in the $L'$-band observed with COMIC/ADONIS+RASOIR and NACO in 1999 and 2002 respectively. The green $\times$ marks the approximate position of Sgr~A*. In 1999 and 2002, the position of S2 and Sgr~A* are confused because of its close proximity to each other. Some re-identified S-stars are marked with a light green circle. In 1999, the orbital spatial position of the S-cluster stars S71 and S72 coincide which results in the bright spot marked with a light green circle. The right handed side shows orbits of the S-stars S33 (marked), S50, S71/S72 (marked), and S87 (marked). The position of S65 can be used for orientation in these plots (please see Fig. \ref{fig:X7_2016_NACO} for comparison). The orbit of S50 is highlighted in red. The empty circle of S33 in 2002 corresponds to the position of the star in the high-pass filtered image.}
\label{fig:comic_1999}
\end{figure*}

In Fig. \ref{fig:comic_1999}, we present the results of the \textbf{long-time} survey of X7 in the $L'$-band with COMIC/ADONIS+RASOIR (1999) in combination with the NACO data (2002, 2003 - 2018 is shown in Fig. \ref{fig:X7_2016_NACO}). 
For the image presented in Fig. \ref{fig:comic_1999} which is observed with COMIC/ADONIS+RASOIR, we use a high pass filter to highlight features of the S-cluster. In both images, we clearly detect the structure of the S-cluster (Fig. \ref{fig:comic_1999}). 
Even though the resulting COMIC/ADONIS+RASOIR image of 1999 suffers from a decreased magnitude sensitivity, we are still able to identify several (isolated) sources including the spherical shaped \textbf{bow shock} source X7 at the position of S50. As indicated by the orbital plots presented on the \textbf{right-handed} side of Fig. \ref{fig:comic_1999}, we identify the nearby S-cluster stars S33, S71/S72, S65, and S87 and mark them accordingly. \textbf{Moreover,} we include the K-band based orbit of S50 (red dot) in the presented COMIC/ADONIS+RASOIR data of 1999 (red ellipse).

\section{Data}
\label{sec:appendix_Data}

Here, we list \textbf{the NACO} and SINFONI data. Parts of this data \textbf{were} analysed in various publications like \cite{muzic2010}, \cite{Witzel2012}, \cite{Eckart2013a}, \cite{Zajacek2014}, \cite{Valencia-S.2015}, \cite{Shahzamanian2016}, \cite{Parsa2017}, and \cite{Peissker2019, Peissker2020a, Peissker2020c, Peissker2020d, Peissker2020b}. These listed publications underline the robustness of the used data. For the sake of completeness, it should be noted that \cite{Parsa2017} derived \textbf{with the} here used data the gravitational \textbf{redshift of} S2 caused by the SMBH. This was later independently confirmed by \cite{gravity2018} and indicates the quality of the \textbf{data reduction} process applied to the data. 

\begin{table*}[htbp!]
        \centering
        \begin{tabular}{cccccc}
        \hline\hline
        \\      Date & Observation ID  & \multicolumn{3}{c}{Amount of on source exposures} & Exp. Time \\  \cline{3-5} &  & Total & Medium & High &  \\
        (YYYY:MM:DD) &  &  &  &  & (s) \\ \hline\hline %table heading
        
        2005.06.16 & 075.B-0547(B) &  20  &  12   &  8  &    300  \\
        2005.06.18 & 075.B-0547(B) &  21  &  2   &  19  &    60  \\
        2006.03.17 & 076.B-0259(B) &  5  &  0   &  3  &    600  \\
        2006.03.20 & 076.B-0259(B) &   1 &   1  &   0 &    600  \\
        2006.03.21 & 076.B-0259(B) &  2  &   2  &  0  &    600  \\
        2006.04.22 & 077.B-0503(B) &  1  &   0  &  0  &    600  \\
        2006.08.17 & 077.B-0503(C) &  1  &   0  &  1  &    600  \\
        2006.08.18 & 077.B-0503(C) &  5  &   0  &  5  &    600  \\
        2006.09.15 & 077.B-0503(C) &  3  &   0  &  3  &    600  \\
        2007.03.26 & 078.B-0520(A) &  8  &   1  &  2  &    600  \\
        2007.04.22 & 179.B-0261(F) &  7  &   2  &  1  &    600 \\
        2007.04.23 & 179.B-0261(F) &  10 &   0  &  0  &    600 \\
        2007.07.22 & 179.B-0261(F) &  3  &   0  &  2  &    600  \\
        2007.07.24 & 179.B-0261(Z) &  7  &   0  &  7  &    600  \\
        2007.09.03 & 179.B-0261(K) & 11  &   1  &  5  &    600  \\
        2007.09.04 & 179.B-0261(K) &  9  &   0  &  0  &    600  \\
        2008.04.06 & 081.B-0568(A) &  16 &   0  &  15   &    600  \\
        2008.04.07 & 081.B-0568(A) &   4 &   0  &   4 &    600  \\
        2009.05.21 & 183.B-0100(B) &  7 &   0  &  7   &    600  \\
        2009.05.22 & 183.B-0100(B) &   4 &   0  &   4 &    400  \\
        2009.05.23 & 183.B-0100(B) &  2 &   0  &  2  &    400  \\
        2009.05.24 & 183.B-0100(B) &  3 &   0  &  3  &    600  \\
        
        \hline  \\
        \end{tabular}   
        \caption{SINFONI data of 2005, 2006, 2007, 2008, and 2009. The total amount of data is listed.}
        \label{tab:data_sinfo1}
        \end{table*}
\begin{table*}[htbp!]
        \centering
        \begin{tabular}{cccccc}
        \hline\hline
        \\      Date & Observation ID  & \multicolumn{3}{c}{Amount of on source exposures} & Exp. Time \\  \cline{3-5} &  & Total & Medium & High &  \\
        (YYYY:MM:DD) &  &  &  &  & (s) \\ \hline\hline %table heading
        
        2010.05.10 & 183.B-0100(O) & 3 &   0  &  3   &    600  \\
        2010.05.11 & 183.B-0100(O) &  5 &   0  &   5 &    600  \\
        2010.05.12 & 183.B-0100(O) & 13 &   0  &  13  &    600  \\
        2011.04.11 & 087.B-0117(I) &  3 &   0  &   3 &    600  \\
        2011.04.27 & 087.B-0117(I) & 10 &   1  &  9  &    600  \\
        2011.05.02 & 087.B-0117(I) & 6 &   0  &  6  &    600  \\
        2011.05.14 & 087.B-0117(I) & 2 &   0  &  2  &    600  \\
        2011.07.27 & 087.B-0117(J)/087.A-0081(B) & 2 &   1  &  1  &    600  \\     
        2012.03.18 & 288.B-5040(A) &  2 &   0  &   2 &    600  \\
        2012.05.05 & 087.B-0117(J) & 3 &   0  &  3  &    600  \\
        2012.05.20 & 087.B-0117(J) & 1 &   0  &  1  &    600  \\
        2012.06.30 & 288.B-5040(A) & 12 &   0  &  10  &    600  \\
        2012.07.01 & 288.B-5040(A) & 4 &   0  &  4  &    600  \\
        2012.07.08 & 288.B-5040(A)/089.B-0162(I)& 13 &   3  &  8  &    600  \\
        2012.09.08 & 087.B-0117(J)  & 2 &   1  &  1  &    600  \\
        2012.09.14 & 087.B-0117(J)  & 2 &   0  &  2  &    600  \\   
        2013.04.05 & 091.B-0088(A)  &  2 &   0  &   2 &    600  \\
        2013.04.06 & 091.B-0088(A)  & 8 &   0  &  8  &    600  \\
        2013.04.07 & 091.B-0088(A)  & 3 &   0  &  3  &    600  \\
        2013.04.08 & 091.B-0088(A)  & 9 &   0  &  6  &    600  \\
        2013.04.09 & 091.B-0088(A)  & 8 &   1  &  7  &    600  \\
        2013.04.10 & 091.B-0088(A)  & 3 &   0  &  3  &    600  \\
        2013.08.28 & 091.B-0088(B)  & 10 &  1  &  6  &    600 \\
        2013.08.29 & 091.B-0088(B)  & 7 &  2   &  4  &    600 \\
        2013.08.30 & 091.B-0088(B)  & 4 &  2   &  0  &    600 \\
        2013.08.31 & 091.B-0088(B)  & 6 &  0   &  4  &    600 \\
        2013.09.23 & 091.B-0086(A)  & 6 &   0  &  0  &    600  \\
        2013.09.25 & 091.B-0086(A)  & 2 &   1  &  0  &    600  \\
        2013.09.26 & 091.B-0086(A)  & 3 &   1  &  1  &    600  \\   
        
        \hline  \\
        \end{tabular}
        
        \caption{SINFONI data of 2010, 2011, 2012, and 2013.}
        \label{tab:data_sinfo2}
        \end{table*}

\begin{table*}[htbp!]
        \centering
        \begin{tabular}{cccccc}
        \hline\hline
        \\      Date & Observation ID  & \multicolumn{3}{c}{Amount of on source exposures} & Exp. Time \\  \cline{3-5} &  & Total & Medium & High &  \\
        (YYYY:MM:DD) &  &  &  &  & (s) \\ \hline\hline %table heading
        
        2014.02.27 & 092.B-0920(A) &  4 &   1  &  3  &    600   \\
        2014.02.28 & 091.B-0183(H) &  7 &   3  &  1  &    400   \\
        2014.03.01 & 091.B-0183(H) & 11 &   2  &  4  &    400  \\
        2014.03.02 & 091.B-0183(H) &  3 &   0  &  0  &    400   \\
        2014.03.11 & 092.B-0920(A) & 11 &   2  &  9  &    400   \\
        2014.03.12 & 092.B-0920(A) & 13 &   8  &  5  &    400   \\
        2014.03.26 & 092.B-0009(C) & 9  &   3  &  5  &    400   \\
        2014.03.27 & 092.B-0009(C) & 18 &   7  &  5  &    400   \\
        2014.04.02 & 093.B-0932(A) & 18 &   6  &  1  &    400    \\
        2014.04.03 & 093.B-0932(A) & 18 &   1  &  17 &    400    \\
        2014.04.04 & 093.B-0932(B) & 21 &   1  &  20 &    400       \\
        2014.04.06 & 093.B-0092(A) &  5 &   2  &  3  &    400      \\
        2014.04.08 & 093.B-0218(A) & 5  &   1  &  0  &    600   \\
        2014.04.09 & 093.B-0218(A) &  6 &   0  &  6  &    600   \\
        2014.04.10 & 093.B-0218(A) & 14 &   4  &  10  &    600   \\
        2014.05.08 & 093.B-0217(F) & 14 &   0  &  14  &    600   \\
        2014.05.09 & 093.B-0218(D) & 18 &   3  &  13  &    600   \\
        2014.06.09 & 093.B-0092(E) &  14 &   3  &  0  &    400   \\
        2014.06.10 & 092.B-0398(A)/093.B-0092(E) & 5 &   4  &  0   & 400/600 \\
        2014.07.08 & 092.B-0398(A)  & 6 &   1  &  3   &    600 \\
        2014.07.13 & 092.B-0398(A)  & 4 &   0  &  2   &    600 \\
        2014.07.18 & 092.B-0398(A)/093.B-0218(D)  & 1 &   0  &  0   &    600 \\
        2014.08.18 & 093.B-0218(D)  & 2 &   0  &  1   &    600 \\ 
        2014.08.26 & 093.B-0092(G)  &  4 &   3  &   0 &    400   \\
        2014.08.31 & 093.B-0218(B)  & 6 &   3   &   1 &    600 \\
        2014.09.07 & 093.B-0092(F)  & 2 &   0  &  0  &    400   \\
        2015.04.12 & 095.B-0036(A)  & 18 &  2 & 0 & 400 \\
        2015.04.13 & 095.B-0036(A)  & 13 &  7 & 0 & 400 \\
        2015.04.14 & 095.B-0036(A)  & 5  &  1 & 0 & 400 \\
        2015.04.15 & 095.B-0036(A)  & 23 &  13  & 10 & 400 \\
        2015.08.01 & 095.B-0036(C)  & 23 &   7  & 8  & 400 \\
        2015.09.05 & 095.B-0036(D)   17 &  11  & 4  & 400 \\

        \hline  \\
        \end{tabular}
        
        \caption{SINFONI data of 2014 and 2015.}
        \label{tab:data_sinfo3}
        \end{table*}
        
\begin{table*}[htbp!]
        \centering
        \begin{tabular}{cccccc}
        \hline\hline
        \\      Date & Observation ID  & \multicolumn{3}{c}{Amount of on source exposures} & Exp. Time \\  \cline{3-5} &  & Total & Medium & High &  \\
        (YYYY:MM:DD) &  &  &  &  & (s) \\ \hline\hline %table heading
        
        2017.03.15 & 598.B-0043(D) &  5 &   2  &  0  &    600   \\
        2017.03.19 & 598.B-0043(D) & 11 &   0  &  5  &    600   \\
        2017.03.20 & 598.B-0043(D) & 15 &   4  &  11 &    600   \\
        2017.03.21 & 598.B-0043(D) & 1  &   0  &  0  &    600  \\
        2017.05.20 & 0101.B-0195(B) & 8 &   2  &  6  &    600   \\
        2017.06.01 & 598.B-0043(E) & 5  &   0  &  3  &    600   \\
        2017.06.02 & 598.B-0043(E) & 8  &   0  &  8  &    600  \\
        2017.06.29 & 598.B-0043(E) & 4  &   2  &  17 &    600   \\
        2017.07.20 & 0101.B-0195(C) & 8 &   5  &  0  &    600   \\
        2017.07.28 & 0101.B-0195(C) & 6 &   0  &  0  &    600   \\
        2017.07.29 & 0101.B-0195(D) & 9 &   0  &  0  &    600   \\
        2017.08.01 & 0101.B-0195(E) & 4 &   0  &  0  &    600   \\
        2017.08.19 & 598.B-0043(F) &  8 &   0  &  2  &    600   \\
        2017.09.13 & 598.B-0043(F) &  8 &   0  &  0  &    600   \\
        2017.09.15 & 598.B-0043(F) & 10  &  1  &  1  &    600   \\
        2017.09.29 & 598.B-0043(F) &  2  &  0  &  0  &    600   \\
        2017.10.15 & 0101.B-0195(F) & 2  &  0  &  0  &    600   \\
        2017.10.17 & 0101.B-0195(F) & 4  &  0  &  0  &    600   \\
        2017.10.23 & 598.B-0043(G) &  3  &  0  &  0  &    600   \\
        \hline  \\
        \end{tabular}
        
        \caption{SINFONI data of 2017.}
        \label{tab:data_sinfo4}
        \end{table*}

\begin{table*}[htbp!]
        \centering
        \begin{tabular}{cccccc}
        \hline\hline
        \\      Date & Observation ID  & \multicolumn{3}{c}{Amount of on source exposures} & Exp. Time \\  \cline{3-5} &  & Total & Medium & High &  \\
        (YYYY:MM:DD) &  &  &  &  & (s) \\ \hline\hline %table heading
        
        2018.02.13 & 299.B-5056(B) &  3 &   0  &  0  &    600   \\
        2018.02.14 & 299.B-5056(B) &  5 &   0  &  0  &    600   \\
        2018.02.15 & 299.B-5056(B) &  5 &   0  &  0  &    600   \\
        2018.02.16 & 299.B-5056(B) &  5 &   0  &  0  &    600   \\
        2018.03.23 & 598.B-0043(D) &  8 &   0  &  8  &    600   \\
        2018.03.24 & 598.B-0043(D) &  7 &   0  &  0  &    600   \\
        2018.03.25 & 598.B-0043(D) &  9 &   0  &  1  &    600   \\
        2018.03.26 & 598.B-0043(D) & 12 &   1  &  9  &    600  \\
        2018.04.09 & 0101.B-0195(B) &  8 &   0  &  4  &    600   \\
        2018.04.28 & 598.B-0043(E) & 10 &   1  &  1  &    600   \\
        2018.04.30 & 598.B-0043(E) & 11 &   1  &  4  &    600  \\
        2018.05.04 & 598.B-0043(E) & 17 &   0  &  17  &    600   \\
        2018.05.15 & 0101.B-0195(C) &  8 &   0  &  0  &    600   \\
        2018.05.17 & 0101.B-0195(C) &  8 &   0  &  4  &    600   \\
        2018.05.20 & 0101.B-0195(D) &  8 &   0  &  4  &    600   \\
        2018.05.28 & 0101.B-0195(E) &  8 &   3  &  1  &    600   \\
        2018.05.28 & 598.B-0043(F) &  4 &   0  &  4  &    600   \\
        2018.05.30 & 598.B-0043(F) &  8 &   5  &  3  &    600   \\
        2018.06.03 & 598.B-0043(F) &  8 &   0  &  8  &    600   \\
        2018.06.07 & 598.B-0043(F) & 14 &   1  &  7  &    600   \\
        2018.06.14 & 0101.B-0195(F) &  4 &   0  &  0  &    600   \\
        2018.06.23 & 0101.B-0195(F) &  8 &   1  &  1  &    600   \\
        2018.06.23 & 598.B-0043(G) &  7 &   2  &  1  &    600   \\
        2018.06.25 & 598.B-0043(G) & 22 &   5  &  7  &    600   \\
        2018.07.02 & 598.B-0043(G) &  3 &   0  &  0  &    600   \\
        2018.07.03 & 598.B-0043(G) & 22 &  12  & 10  &    600   \\
        2018.07.09 & 0101.B-0195(G) &  8 &   3  &  1  &    600   \\
        2018.07.24 & 598.B-0043(H) &  3 &   0  &  0  &    600   \\
        2018.07.28 & 598.B-0043(H) &  8 &   0  &  3  &    600   \\
        2018.08.03 & 598.B-0043(H) &  8 &   0  &  1  &    600   \\
        2018.08.06 & 598.B-0043(H) &  8 &   1  &  1  &    600   \\
        2018.08.19 & 598.B-0043(I) & 12 &   2  & 10  &    600   \\
        2018.08.20 & 598.B-0043(I) & 12 &   0  & 12  &    600   \\
        2018.09.03 & 598.B-0043(I) &  1 &   0  &  0  &    600   \\
        2018.09.27 & 598.B-0043(J) & 10 &   0  &  0  &    600   \\
        2018.09.28 & 598.B-0043(J) & 10 &   0  &  0  &    600   \\
        2018.09.29 & 598.B-0043(J) &  8 &   0  &  0  &    600   \\
        2018.10.16 & 2102.B-5003(A) &  3 &   0  &  0  &    600   \\

        \hline  \\
        \end{tabular}
        
        \caption{SINFONI data of 2018.}
        \label{tab:data_sinfo5}
        \end{table*}

\begin{table*}[h!]
\centering
\begin{tabular}{ccccc}
\hline
\hline
\multicolumn{5}{c}{NACO}\\
\hline
Date  & Observation ID & \multicolumn{1}{p{1.5cm}}{\centering number \\ of exposures }   & \multicolumn{1}{p{1.5cm}}{\centering Total \\ exposure time(s) } & $\lambda$ \\
\hline
2002.07.31 & 60.A-9026(A)  & 61  & 915     & K \\
2003.06.13 & 713-0078(A)   & 253 & 276.64  & K \\
2004.07.06 & 073.B-0775(A) & 344 & 308.04  & K \\
2004.07.08 & 073.B-0775(A) & 285 & 255.82  & K \\
2005.07.25 & 271.B-5019(A) & 330 & 343.76  & K \\
2005.07.27 & 075.B-0093(C) & 158 & 291.09  & K \\
2005.07.29 & 075.B-0093(C) & 101 & 151.74  & K \\
2005.07.30 & 075.B-0093(C) & 187 & 254.07  & K \\
2005.07.30 & 075.B-0093(C) & 266 & 468.50  & K \\
2005.08.02 & 075.B-0093(C) & 80  & 155.77  & K \\
2006.08.02 & 077.B-0014(D) & 48  & 55.36   & K \\
2006.09.23 & 077.B-0014(F) & 48  & 55.15   & K \\
2006.09.24 & 077.B-0014(F) & 53  & 65.10   & K \\
2006.10.03 & 077.B-0014(F) & 48  & 53.84   & K \\
2006.10.20 & 078.B-0136(A) & 47  & 42.79   & K \\
2007.03.04 & 078.B-0136(B) & 48  & 39.86   & K \\
2007.03.20 & 078.B-0136(B) & 96  & 76.19   & K \\
2007.04.04 & 179.B-0261(A) & 63  & 49.87   & K \\ 
2007.05.15 & 079.B-0018(A) & 116 & 181.88  & K \\ 
2008.02.23 & 179.B-0261(L) & 72  & 86.11   & K \\ 
2008.03.13 & 179.B-0261(L) & 96  & 71.49   & K \\ 
2008.04.08 & 179.B-0261(M) & 96  & 71.98   & K \\ 
2009.04.21 & 178.B-0261(W) & 96  & 74.19   & K \\ 
2009.05.03 & 183.B-0100(G) & 144 & 121.73  & K \\ 
2009.05.16 & 183.B-0100(G) & 78  & 82.80   & K \\ 
2009.07.03 & 183.B-0100(D) & 80  & 63.71   & K \\ 
2009.07.04 & 183.B-0100(D) & 80  & 69.72   & K \\ 
2009.07.05 & 183.B-0100(D) & 139 & 110.40  & K \\ 
2009.07.05 & 183.B-0100(D) & 224 & 144.77  & K \\ 
2009.07.06 & 183.B-0100(D) & 56  & 53.81   & K \\ 
2009.07.06 & 183.B-0100(D) & 104 & 72.55   & K \\ 
2009.08.10 & 183.B-0100(I) & 62  & 48.11   & K \\ 
2009.08.12 & 183.B-0100(I) & 101 & 77.32   & K \\
2010.03.29 & 183.B-0100(L) & 96  & 74.13   & K \\ 
2010.05.09 & 183.B-0100(T) & 12  & 16.63   & K \\ 
2010.05.09 & 183.B-0100(T) & 24  & 42.13   & K \\ 
2010.06.12 & 183.B-0100(T) & 24  & 47.45   & K \\ 
2010.06.16 & 183.B-0100(U) & 48  & 97.78   & K \\
2011.05.27 & 087.B-0017(A) & 305 & 4575    & K \\
2012.05.17 & 089.B-0145(A) & 169 & 2525    & K \\
2013.06.28 & 091.B-0183(A) & 112 & 1680    & K \\
2017.06.16 & 598.B-0043(L) & 36  & 144     & K \\
2018.04.24 & 101.B-0052(B) & 120 & 1200    & K \\
\hline  
\end{tabular}
\caption{K-band data observed with NACO between 2002 and 2018.}
\label{tab:naco_data2}
\end{table*}

\begin{table*}[h!]
\centering
%\begin{tabular}{ccccc}
%\hline
%\hline
%\multicolumn{5}{c}{NACO}\\
%\hline
%Date (UT) & Observation ID & \multicolumn{1}{p{1.5cm}}{\centering number \\ of exposures }   & %\multicolumn{1}{p{1.5cm}}{\centering Total \\ exposure time(s) } & $\lambda$ \\
%\hline

\begin{tabular}{cccc}
\hline
\hline
\multicolumn{4}{c}{NACO}\\
\hline
Date  & Observation ID & \multicolumn{1}{p{1.5cm}}{\centering number \\ of exposures }    & $\lambda$ \\
\hline
2002.08.30 &  060.A-9026(A) & 80  & $L'$\\  % 16 & $L'$ \\
2003.05.10 &  071.B-0077(A) & 56  & $L'$\\  % 16 & $L'$ \\
2004.07.06 &  073.B-0775(A) & 217 & $L'$\\  % 43.3 & $L'$ \\
2005.05.13 &  073.B-0085(E) & 108 & $L'$\\  % 21.6 & $L'$ \\
2005.06.20 &  073.B-0085(F) & 100 & $L'$\\  % 20   & $L'$ \\
2006.05.28 &  077.B-0552(A) &  46 & $L'$\\  %  9.2 & $L'$ \\
2006.06.01 &  077.B-0552(A) & 244 & $L'$\\  % 48.8 & $L'$ \\
2007.03.17 &  078.B-0136(B) &  78 & $L'$\\  % 15.6 & $L'$ \\
2007.04.01 &  179.B-0261(A) &  96 & $L'$\\  % 19.2 & $L'$ \\
2007.04.02 &  179.B-0261(A) & 150 & $L'$\\  % 30   & $L'$ \\
2007.04.02 &  179.B-0261(A) &  72 & $L'$\\  % 14.4 & $L'$ \\
2007.04.06 &  179.B-0261(A) & 175 & $L'$\\  % 35   & $L'$ \\
2007.06.09 &  179.B-0261(H) &  40 & $L'$\\  %  8   & $L'$ \\
2008.05.28 &  081.B-0648(A) &  58 & $L'$\\  % 11.6 & $L'$ \\
2008.08.05 &  179.B-0261(N) &  64 & $L'$\\  % 12.8 & $L'$ \\
2008.09.14 &  179.B-0261(U) &  49 & $L'$\\  %  9.8 & $L'$ \\
2009.03.29 &  179.B-0261(X) &  32 & $L'$\\  %  6.4 & $L'$ \\
2009.03.31 &  179.B-0261(X) &  32 & $L'$\\  %  6.4 & $L'$ \\
2009.04.03 &  082.B-0952(A) &  42 & $L'$\\  %  8.4 & $L'$ \\
2009.04.05 &  082.B-0952(A) &  12 & $L'$\\  %  2.4 & $L'$ \\
2009.09.19 &  183.B-0100(J) & 132 & $L'$\\  % 26.4 & $L'$ \\
2009.09.20 &  183.B-0100(J) &  80 & $L'$\\  % 16   & $L'$ \\
2010.07.02 &  183.B-0100(Q) & 485 & $L'$\\  % 97   & $L'$ \\
2011.05.25 &  087.B-0017(A) & 29  & $L'$\\  %  5.8 & $L'$ \\
2012.05.16 &  089.B-0145(A) & 30  & $L'$\\  %  6   & $L'$ \\
2013.05.09 &  091.C-0159(A) & 30  & $L'$\\  %  6   & $L'$ \\
2016.03.23 &  096.B-0174(A) & 60  & $L'$\\  %  12  & $L'$ \\
2017.03.23 &  098.B-0214(B) & 30  & $L'$\\  %   6  & $L'$ \\
2018.04.22 & 0101.B-0065(A) & 68  & $L'$\\  % 13.6 & $L'$ \\
2018.04.24 & 0101.B-0065(A) & 50  & $L'$\\  % 10   & $L'$ \\
\hline  
\end{tabular}
\caption{$L'$-band data observed with NACO between 2002 and 2018.}
\label{tab:naco_data3}
\end{table*}

%%%% End of aa.dem
\end{document}